\documentclass[letter, 12pt]{article}
\usepackage[american]{babel}
\usepackage{blindtext}
\usepackage[margin=2.54cm]{geometry} 
\usepackage[utf8]{inputenc} 
\usepackage[hidelinks]{hyperref}
\hypersetup{colorlinks=true, linkcolor=red, filecolor=magenta, urlcolor=cyan, citecolor=blue} 
\usepackage{verbatim} 
\usepackage{enumerate}
\usepackage{lscape} 
\usepackage{authblk} 

\usepackage{caption} 
\usepackage{chngcntr} 

\usepackage{indentfirst} 
\usepackage[hang,flushmargin]{footmisc} 
\usepackage{setspace}
\usepackage{booktabs} 
\usepackage{array} 
\usepackage{multirow} 
\usepackage[table,xcdraw]{xcolor}
\usepackage[para,online,flushleft]{threeparttable}
\usepackage{tabularx}

\usepackage{subcaption} 
\usepackage{graphicx} 
\usepackage{float} 
\usepackage{epstopdf} 
\usepackage{tikz}
\usetikzlibrary{decorations.pathreplacing}
\usetikzlibrary{shapes,arrows}

\usepackage{apacite} 
\usepackage{natbib} 

\usepackage{amsmath} 
\usepackage{amssymb} 
\usepackage{pifont} 
\usepackage{relsize} 
\usepackage{nccmath} 
\usepackage{mathtools} 
\usepackage{array} 

\usepackage{amsfonts}

\usepackage{tabularx}

\begin{document}

	\title{\Large The Effect of Financial Resources on Fertility: Evidence from Administrative Data on Lottery Winners}

	\author{\normalsize Yung-Yu Tsai\thanks{Truman School of Government and Public Affairs, University of Missouri; {\scriptsize Email: ytsai@mail.missouri.edu}}, Hsing-Wen Han\thanks{Department of Accounting, Tamkang University; Email: hwhan466@mail.tku.edu.tw}, Kuang-Ta Lo\thanks{Department of Public Finance, National Chengchi University; Email: vancelo@nccu.edu.tw}, Tzu-Ting Yang\thanks{Institute of Economics, Academia Sinica, Email: ttyang@econ.sinica.edu.tw.} \medskip}
	
	\maketitle
	
	\begin{abstract}
This paper utilizes wealth shocks from winning lottery prizes to examine the causal effect of financial resources on fertility. We employ extensive panels of administrative data encompassing over 0.4 million lottery winners in Taiwan and implement a triple-differences design. Our analyses reveal that a substantial lottery win can significantly increase fertility, the implied wealth elasticity of which is around 0.06. Moreover, the primary channel through which fertility increases is by prompting first births among previously childless individuals. Finally, our analysis reveals that approximately 25\% of the total fertility effect stems from increased marriage rates following a lottery win.
	\end{abstract}

	\enlargethispage{2\baselineskip}%
	
	\thispagestyle{empty}\newpage \setcounter{page}{1}\baselineskip=24pt

	\section{Introduction}
	
	Over the past fifty years, most countries worldwide have witnessed a significant decline in fertility rates \citep{oecd2019fertility}. This trend has sparked concerns about an aging population, shortages in workforce participation, and a decrease in tax revenue \citep{bloom2010cost,caldwell2006policy,sleebos2003low}. As a response, many countries have implemented programs that provide financial incentives for having children, aiming to stimulate the birth rate. On average, public spending on these pro-natality incentives accounts for 1.1\% of GDP in OECD countries \citep{oecd2019family}. The rationale behind these policies lies in the premise that people often do not accumulate sufficient wealth to afford the costs of raising children. Therefore, the enhancement of individual financial resources should theoretically increase the desire for more children.\footnote{How do people's financial resources affect their fertility behaviors? This issue has been discussed in social science for a long time, dating back to Malthus's influential book published in the 18th century \citep{Malthus1798essay}.} A seminal paper by \citet{becker1960economic} incorporated fertility decisions into an economic model, with his framework suggesting that the demand for children could be viewed as a demand for durable goods that have few substitutes. According to consumer theory, goods with few substitutes are essentially ``normal.'' Therefore, the theory predicts that an increase in the lifetime income of individuals should encourage greater desired fertility. However, cross-sectional evidence largely contradicts this theoretical prediction, suggesting instead a negative relationship between income and fertility \citep{jones2008fertility, gauthier2007impact,jones2008chapter}.\footnote{Figure \ref{fig.gdp_fertility} of the Online Appendix demonstrates that countries with higher per capita income tend to have lower total fertility levels.}

	The gap between theory and cross-sectional evidence is due to the fact that identifying the causal effect of lifetime income on fertility suffers from various challenges. First, it is possible that the observed relationship between these two factors reveals a reverse causality; for example, recent literature suggests that the arrival of children can substantially reduce a mother's income \citep{kleven2019children,sieppi2019parenthood,cortes2020children,de2020child,berniell2021gender}. Second, an individual's income and fertility are usually jointly determined \citep{francesconi2002joint,del2002effect,huttunen2016effect}. Both working and raising children are time-consuming activities, and so a sudden increase in wages could increase the relative price of having children. Thus, the income effect always confounds with the substitution effect when using labor earnings as a proxy for lifetime income.\footnote{ In order to overcome these difficulties, several recent studies have used arguably exogenous variations in income or wealth, such as male job loss or home equity increases, to identify the effect of financial resources on fertility \citep{ager2019structural, alam2018income,huttunen2016effect,lovenheim2013family,black2013children,lindo2010children}. All of these studies provide evidence that financial resources can indeed positively affect fertility. However, these events could still confound with other factors that influence an individual's fertility decision. For example, male income loss might change the division of labor between a husband and wife, which in turn might alter the decision to have a child. House price appreciation, for its part, could have a positive impact on the local economy and government revenue, which may then affect fertility through other channels (e.g., an increase in child-related cash transfers).}
	
	The ideal way to estimate the causal impact of financial resources is based on the fertility responses to randomly assigned changes in the lifetime income of individuals. Following this idea, we examine the fertility impact of large and unexpected non-labor income shocks induced by winning a lottery prize. Specifically, we study this issue by leveraging long panels of administrative data on more than 0.4 million lottery winners in Taiwan. The data allow us to track the same individuals over a period of 10 years so that we are able to investigate the effects of cash windfalls on completed fertility. Utilizing a triple-differences design, we compare trends in outcomes for current and future winners who won larger or smaller prizes.\footnote{Future winners are individuals who win large or small lottery prizes in their later years, so their current outcomes cannot be influenced by these future wins. Using future winners helps account for potential unobserved differences between individuals who tend to win smaller versus larger lottery amounts. Since these future winners do not actually win during the estimation period, we assign them a ``placebo'' winning year by subtracting 6 from their actual first win year. The future winners, therefore, serve as a control group by providing counter-factual trends in outcomes had they not won the lottery. Comparing the current outcomes of actual winners to these future winners who have not yet won allows us to isolate the causal effect of the lottery prize amount. Details on these future winners are provided in Section \ref{sec:em}.}
	
We obtain three key findings. First, our results show that a 10 million NT\$ ( $\approx$ 330,000 US\$) windfall can significantly increase the number of children individuals have by 0.05. Specifically, for every 100 individuals affected by winning a lottery prize, five more children are born to them within 6 years compared to what would have happened without winning the prize. This corresponds to a 15\% increase relative to the comparison group of future winners, who have, on average, 0.32 more children during the same 6-year period after a ``placebo'' lottery-winning year.\footnote{In other words, for every 100 individuals who did not win a prize during the current time, 32 children were born by the sixth year after a ``placebo" lottery-winning year.} The implied wealth elasticity of fertility is around 0.06, which is at the lower end of estimates from previous research examining other resource shocks \citep{ager2019structural, alam2018income,huttunen2016effect,lovenheim2013family,black2013children,lindo2010children}.\footnote{For example, \citet{lovenheim2013family} find wealth elasticity in relation to fertility of 0.13.}

Furthermore, we find positive fertility responses for both young and middle-aged individuals. Given that the average age of the middle-aged sample is around 43 by the end of the observation period, this suggests our results reveal increases not only due to the timing of fertility being brought forward, but also due to a rise in total fertility. Finally, the fertility effect is stronger for those receiving larger windfall gains or having lower pre-existing wealth levels. This implies that individuals decide to have children only after they have accumulated sufficient resources to cover the costs of raising children.

	Second, large cash windfalls increase fertility primarily by inducing childless individuals to have their first child (the extensive margin). In contrast, lottery wins have a negligible impact on subsequent births for those who already have children (the intensive margin), which aligns with \citet{becker1960economic}'s supposition that income elasticity for the quantity of children should be small when parents trade-off between child quality and quantity. To test this notion, we restricted the sample to parents and used children's college attendance as a proxy for quality. The results show that lottery prizes do not affect overall college attendance but significantly increase overseas study, which involves higher cost and is seen as higher quality. This implies that parents opt to invest their money into child quality rather than having more children.
	
	Lastly, given that parenthood is associated with marital status, we investigated heterogeneous effects by marital status---single or married. Subgroup analysis suggests that the fertility effect is primarily driven by unmarried individuals. Examining impacts on marriage provides insights into this mechanism, and we found that a 10 million NT\$ windfall increases marriage rates by 2 percentage points. To quantify how much of the fertility response is explained by changes in marriage, we implemented a causal mediation analysis \citep{hsia2021causal,breivik2022career}, which decomposes the total effect into direct and indirect components via the intermediate effect on marriage. Our results reveal that approximately one-fourth of the overall fertility effect can be attributed to increased marriage rates. These findings shed light on a mechanism whereby windfalls influence fertility decisions in part by making people more likely to get married, as theory suggests \citep{Malthus1798essay,becker1960economic,ahn2002note}. Cash infusions help meet the financial prerequisites for family formation, both in terms of marriage and childbearing, which demonstrates the important interplay between wealth, marriage, and fertility.
		
	Our paper makes several contributions to the existing literature. To begin with, a lottery-induced windfall provides an unexpected, random, and substantial change in lifetime income that is salient to individuals and does not cofound with other factors that might affect fertility.  
	Furthermore, Taiwan's high lottery participation rate allows us to analyze a more general population.\footnote{This paper includes two primary types of lotteries in Taiwan. The first is the Public Welfare Lottery. According to the Social Intention Survey of Public Welfare Lotteries \citep{hsiao2013} conducted by Academia Sinica, approximately 68\% of individuals aged 18 and above have purchased Public Welfare Lottery tickets at least once. The second type is the Taiwan Receipt Lottery. Basically, any person who purchases any goods or services and receives an invoice has ``participated'' in the lottery. Furthermore, according to government statistics, about 70\% of the winning invoices have been redeemed \citep{fia2023a,fia2023b}, implying that the majority of the population keeps the invoices and regularly matches them with the announced winning numbers.}
	Prior work has focused on specific groups like job-losers \citep{huttunen2016effect,lindo2010children}, homeowners \citep{lovenheim2013family}, farmers \citep{ager2019structural, alam2018income}, or miners \citep{black2013children}. Our estimates thus offer greater external validity and generalizability. For instance, studies using a husband's income loss are limited to married couples, whereas our lottery variation includes unmarried individuals, too. We find effects operating through marital decisions and single individuals' fertility.
		
	To our knowledge, we are among the first to leverage lottery-induced wealth shocks to study the impact of financial resources on fertility. We find a positive effect and investigate potential mechanisms, thereby adding to limited existing work.\footnote{The previous literature has studied the effects of lottery wins on labor supply \citep{imbens2001estimating,furaker2009gambling,cesarini2017effect,picchio2018labour}, marriage decision \citep{hankins2011lucky}, individual bankruptcy \citep{hankins2011ticket}, consumption \citep{kuhn2011effects}, stock participation \citep{cesarinistock}, college attendance \citep{bulman2021parental}, and health \citep{apouey2015winning,lindahl2005estimating,kim2021effects,lindqvist2020long}.} Two contemporaneous studies using the design of lottery wins from the US \citep{bulman2022effect} and Sweden \citep{cesarini2023fortunate} provide the most relevant comparisons to our work.\footnote{\citet{bulman2022effect} examines the impacts of wealth shocks on home ownership, marriage, and fertility, with a focus on analyzing the effect on home ownership. \citet{cesarini2023fortunate} investigates the impacts of windfall gains on marriage and fertility but concentrates more on examining the influence on marriage.} \citet{bulman2022effect} use a similar tax data and identification strategy, including future winners for comparison. However, they find little fertility effect from wealth shocks. In contrast, \citet{cesarini2023fortunate} demonstrate that Swedish lottery wins of approximately 100,000 US\$ significantly increase cumulative fertility by 0.021 children within five years, thereby aligning with our estimates. Our analysis builds on this nascent lottery literature by investigating potential mechanisms that drive fertility findings. We explore the lottery's impact on parents' quantity versus quality trade-off., with the results revealing that winners might invest more in child quality through overseas education rather than having additional children. We also conduct causal mediation analysis to quantify marriage's role in the fertility effect.

Finally, there is limited empirical evidence regarding the heterogeneous effects of resource shocks on fertility according to individual socioeconomic backgrounds. Our administrative data contain a wide range of lottery win amounts and detailed personal information. This feature helps us enrich the existing literature by painting a richer picture of how wealth/income affects people's fertility decisions.

	The remainder of this paper is organized as follows. In Section \ref{sec:data}, we discuss our data and the sample selection process. Section \ref{sec:em} presents our empirical strategy, whilst in Section \ref{sec:result}, we present the main results and carry out robustness checks. Section \ref{sec:other_o} illustrates the effect of cash windfalls on other related outcomes. Section \ref{sec:discussion} compares our results with the findings from previous studies, whilst section \ref{sec:conclusion} provides concluding remarks and some future research recommendations.

\section{Data and Sample}\label{sec:data}
	
\subsection{Data}

	We base our analysis on several administrative records: 
	1) Income registry file
	2) Wealth registry file 
	3) Household registration file, 
	and 4) College enrollment file, provided by Taiwan's Fiscal Information Agency (FIA). 
	All files contain individual identifiers (i.e., scrambled personal ID), which allows us to merge them at the individual level.  
	
	Our lottery data information mainly comes from the income statement file, which contains each payment made to an individual on an annual basis. Some are based on third-party reported income sources, such as wage income, interest income, pension income, and lottery income. The remaining records are from self-reported information, namely, rental income, business income, and agricultural income. The records cover all lottery winners who won more than 2,000 NT\$ (about 66 US\$), because only lottery prizes above this amount are subject to a tax rate of 20\% and reported to the FIA. During our sample period, three main types of lottery games were run by the Taiwanese government, namely, (1) Public Welfare Lottery, (2) Taiwan Receipt Lottery, and (3) Taiwan Sports Lottery. We exclude Sports Lottery winners from our samples, as they do not win the prize by ``chance.'' Instead, their winnings could be due to their experience and professional judgment.\footnote{Taiwan Sports Lottery was initiated in 2008. It is the only legal form of sports gambling in Taiwan and includes national and international sports, such as those run by MLB, NBA, and FIFA. The rules of the game include placing bets on winners/losers, total scores, score gaps, etc.} In the Online Appendix \ref{app: add_lottery}, we briefly introduce the Public Welfare Lottery and the Taiwan Receipt Lottery.
	
	The income registry file includes the following information: 1) Taxpayer ID (i.e., the winner); 2) lottery prize amount; and 3) the bank's ID where the prize is redeemed. Since each lottery game uses specific banks for prize redemption, we can use bank ID to exclude Sports Lottery winners. We sum the prizes won by individuals on a yearly basis in order to calculate annual lottery income.\footnote{We only have information on the total amount of lottery prizes at specific redeemed agencies for each individual within a year. If a person redeems multiple tickets at a single agency, we do not have detailed information on each ticket.} 
		
	To measure individuals' financial resources, following \citet{lien2021wealth} and \citet{chu2019variations}, we utilize the income registry file and wealth registry file to construct individual-level wealth data. The details for constructing this wealth dataset are discussed in the Online Appendix \ref{app: w_data}. The data on demographics originate from the household registration file, which is yearly-based and contains the individual's gender, year of birth, location of birth, place of residence, year of marriage, spouse's ID, and parents' IDs (father and mother). We use birth year and parents' IDs to construct the fertility outcome measure, i.e., the number of children that an individual has in a given year.\footnote{We have the record on birth father and mother, regardless of whether the parents are married when the children were born, as long as the biological father is identified and reported to the government. We do not have information on the biological father of an unmarried mother's child if the father was absent and unknown when at the child's birth.} Using the marriage year and spouse's ID, we obtain the marital outcome of whether an individual has ever been married.
	
	The college enrollment file includes two sources: 1) third-party reported college enrollment records from all domestic colleges in Taiwan and 2) self-reported college enrollment as listed as a deduction item on income tax returns.\footnote{College tuition payment for dependents is categorized as a special deduction item. That is, regardless of whether the taxpayers choose the standard deduction or itemized deductions, they can still list tuition costs as an extra deduction from their income.} The second source includes both college attendance at domestic and overseas colleges. We use these data to define college attendance outcomes. The third-party reported data on domestic college enrollment are more comprehensive and cover the whole population, including those who do not file a tax return. However, self-reporting on college enrollment, which is our primary source for overseas study, only covers people who file a tax return. Based on our definition, around 1.5\% of college-aged people in our sample are categorized as studying abroad, which is quite close to the government statistics.\footnote{According to statistics from the Ministry of Education (MOE), Taiwan, around 57 thousand students are currently studying abroad, accounting for roughly 1.5\% of the population of college students.} 
	
\subsection{Sample}\label{sec: sample}

	We impose several restrictions to construct the estimation sample. First, individuals must be aged 20-44 at the time of winning---the primary childbearing years. Second, we limit the sample to those who first won lottery prizes of at least 5,000 NT\$ in the study period.\footnote{Due to data limitations, we cannot observe income records before 2004.} This restriction makes our estimation sample representative of the broader population, based on observable characteristics, and allows us to control for previous lottery winning amounts. In robustness checks, we eliminate this restriction, using a minimum win of 2,000 NT\$, the smallest observable prize. Third, since the number of children people can biologically have is bounded from above,\footnote{Childbearing takes almost one year (i.e., ten months), and in most cases, only one child is born each time. Thus,  the number of births an individual can have within a defined time period is inherently limited.} for sufficiently large windfall gains, the marginal effect on fertility of an additional dollar of lottery prize is likely to be very small. Therefore, following a previous study by \citet{picchio2018labour}, we exclude winners who won an extremely large prize above 100 million NT\$. In robustness checks, we use alternative cut-offs of 80 million and 120 million NT\$ for the maximum prize amount. Fourth, we exclude individuals who died during the study period, thus creating a balanced panel. Finally, we track these individuals over 10 years, from 3 years before to 6 years after winning. The sample period is from 2004 to 2018. The final sample contains over 406,963 prize winners across a wide range of windfall amounts. Table \ref{tab.distribution} in the Online Appendix \ref{app: add_t_f} displays the distribution of lottery prizes. The amount of lottery wins is on a post-tax basis and adjusted to 2016 NT\$ using the Consumer Price Index (CPI). 
	 
	Table \ref{tab.descriptive} compares the characteristics of the lottery winners estimation sample to the Taiwanese population aged 20--44 during the sample period.\footnote{We utilize the all individuals aged 20-44 from 2007-2012 to construct population data. The sample size is around 11 million observations. For each individual, we randomly assign one year between 2007-2012 as a placebo "winning year." We then use their individual characteristics from the year prior to this randomly assigned placebo winning year in our analysis.} These characteristics are measured in the year before the lottery win, and all monetary values are adjusted to 2016 NT\$ using the CPI.\footnote{The population is aged 20-44 years old during 2007-2012} Overall, winners largely mirror major demographic attributes of the general public, albeit they are slightly older than the population. Consistent with this fact, a higher proportion are married (46\% vs 41\%) and have more children on average (0.88 vs 0.82). The average annual income of winners is not significantly different from that of the general population. Additionally, the average annual earnings of winners are only slightly higher (by 4,000 NT\$ or 1.4\%) than the general population. Winners are also less wealthy than the general population; however, these differences are minor. In a robustness check, we re-weight the sample to align these characteristics with those of the overall Taiwanese population and demonstrate that our main estimate remains robust despite this issue. 

\section{Empirical Strategy}\label{sec:em}

	In this section, we introduce our empirical strategies that establish causal inferences about how the receipt of lottery prizes affects people's fertility behaviors. First, following the previous literature \citep{cesarini2016wealth,cesarini2017effect,cesarinistock,picchio2018labour}, our specification exploits variations in the size of the lottery win by using prize amounts to measure treatment intensity. This helps us facilitate the interpretation of our findings in terms of the dollar value of the lottery winnings. In addition, we follow the same people over time and investigate their behaviors before and after the year of the lottery win. Therefore, one possible strategy is a difference-in-differences (DID) design, which examines whether people who won larger prizes (first difference) increased fertility after the lottery-winning year (second difference).

    However, a design relying solely on variations in lottery prize amounts could exhibit bias if individuals winning larger and smaller prizes differ in terms of unobservable factors associated with changes in outcomes. Table \ref{tab.balance} in the Online Appendix \ref{app: add_t_f} examines the relationship between prize amounts and pre-lottery characteristics. In Panel A, it is evident that the lottery prize amount is correlated with several winner's traits, such as gender, marital status, and previous winning amounts.\footnote{The panel regresses the winner's characteristics on prize amounts, controlling for the age-fixed effect.} The pattern implies that the amount of lottery winnings may not be entirely random.
    
    To address this concern, inspired by \citet{golosov2021americans}, we further utilize variations in the timing of lottery wins and employ a control group comprised of individuals who first won large or small lottery prizes in later years. Thus, their current outcomes cannot be influenced by lottery wins, which helps account for potential unobserved differences between individuals who tend to win smaller and larger prizes. This empirical strategy is essentially a triple-differences (DDD) design that hinges on three variations in 1) the amount of prizes; 2) observation times (pre- and post-winning); and 3) the timing of the lottery win. Specifically, we compare temporal changes in outcomes for current and future winners who win smaller or larger prizes, following which we estimate the following regression.
    
    \begin{align}\label{child_event}
    	B_{it}&= \alpha_{0} Prize_{i}  + \sum_{s\neq-1} \kappa_{s} \cdot \mathbf{I}[t=L_{i}+s]    +\sum_{s\neq-1}  \lambda_{s} \cdot Prize_{i} \times \mathbf{I}[t=L_{i}+s] \nonumber   \\ 
    	&+ (\alpha_{1} + \alpha_{2} Prize_{i}  + \sum_{s\neq-1} \beta_{s} \cdot \mathbf{I}[t=L_{i}+s]    +\sum_{s\neq-1}  \gamma_{s} \cdot Prize_{i} \times \mathbf{I}[t=L_{i}+s] ) \times Current_{i} \nonumber   \\
    	& + a_{it} + \theta_{t}  + \boldsymbol{X_{i}} \psi  + \varepsilon_{it} 
    \end{align}

\noindent The outcome of interest, represented by $B_{it}$, is the cumulative number of children that an individual $i$ ever has at time t. $Prize_{i}$ denotes the amount of individual $i$'s first lottery win, measured in units of 10 million NT\$ ($\approx$ 330,000 US\$). Event time dummies $I[t=L_{i}+s]$ indicate observations before or after lottery wins, where $L_{i}$ is the year of individual $i$'s first lottery win. Thus, $I[t=L_{i}+s]$ represents an indicator for being $s$ years away from the win, with $s=-3,-2,0,1,2,3,4,5,6$. For instance, $I[t=L_{i}+1]$ is a dummy for the first year after the lottery-winning year. Our sample comprises a balanced panel of individuals observed annually from 3 years ($s=-3$) pre-winning to 6 years ($s=6$) post-winning. We normalize the event time dummy coefficients at the baseline year $s=-1$ to zero. 

This specification includes $Current_{i}$, a dummy variable indicating that an individual $i$ is either a current winner who first won lottery prizes in year $L_{i}$ ($Current_{i}=1$) or a future winner whose first lottery winning year is after $L_{i}+6$ ($Current_{i}=0$). We fully interact $Current_{i}$ with prize amount $Prize_{i}$ and event time dummies $I[t=L_{i}+s]$. For future winners, $L_{i}$ is a ``placebo" winning year determined by subtracting 6 from their actual first winning year. The key identification variables in the regression (\ref{child_event}) are the following third-level interactions: event time dummies $I[t=L_{i}+s]$ interacted with current winner dummy $Current_{i}$ and prize amount $Prize_{i}$. Its coefficients $\gamma_{s}$ measure the effect of a 10 million NT\$ windfall on the outcome of interest.

Since age is a key determinant of an individual's fertility behavior, all specifications include a winner's age-fixed effects $a_{it}$, to control non-parametrically for underlying life-cycle fertility trends.\footnote{Since a female's age relates more directly to fertility, robustness checks add controls for the female household member's age if the winner is male.} Year fixed effects $\theta_t$ capture macroeconomic impacts and general fertility patterns in Taiwan. We also incorporate pre-determined covariates $\boldsymbol{X}_{i}$ measured right before a lottery-winning year, such as the winner's residence, employment status, and financial resources.\footnote{We include the following variables: a set of dummy variables indicating cities/counties of residence, a dummy variable indicating whether the winner was married, a dummy variable indicating whether the winner or their spouse was employed, average household earnings per capita (evenly divided between spouses if married), average household income per capita (evenly divided between spouses if married), average household wealth per capita (evenly divided between spouses if married), the number of cumulative children in the year immediately prior to the lottery winning year, and the lottery winning amount from one, two, and three years before the current lottery winning year. Note that we define the lottery winning year as the year when the winner first won a lottery prize of at least 5,000 NT\$. Therefore, winners could have won another prize below 5,000 NT\$ before the current winning year we use to define treatment.} The error term is represented by $\varepsilon_{it}$. As we follow individuals over time, standard errors in all regressions are clustered at the individual level to account for potential serial correlation.
		
\section{The Effect of Cash Windfalls on Fertility}\label{sec:result}
	
\subsection{Graphical Evidence}

This section illustrates graphically variation sources identifying the causal effect of lottery wins on fertility. Our DDD design essentially compares the fertility trends between current winners and future winners of similar prize amounts. Figure \ref{fig.trend} displays the evolution of total children ever born for current winners who won over 1 million NT\$ (solid line, circle symbol) versus future winners who won over 1 million NT\$ (dashed line, square symbol), from 3 years before to 6 years after winning. Note that the future winners did not receive lottery prizes currently, and their winning year is a ``placebo'' year. The vertical axis shows the outcomes at event time $s$ relative to the base year (s=-1) for each group. 

Three key insights emerge in this regard. First, pre-winning fertility trends are nearly identical between the two groups. Second, an immediate divergence in trends is observed after current winners receive a significant financial windfall, since future winners had not yet won prizes. Third, the effect persists for at least 6 years after winning.\footnote{Divergence is observed at one year after a lottery win (s=1) because pregnancy takes around 10 months.}

\subsection{Main Results}

\begin{figure}[H]
	\centering
	\caption{Trend in the Cumulative Number of Children: Current vs. Future Winners} \label{fig.trend}	 

		\includegraphics[width=1\textwidth]{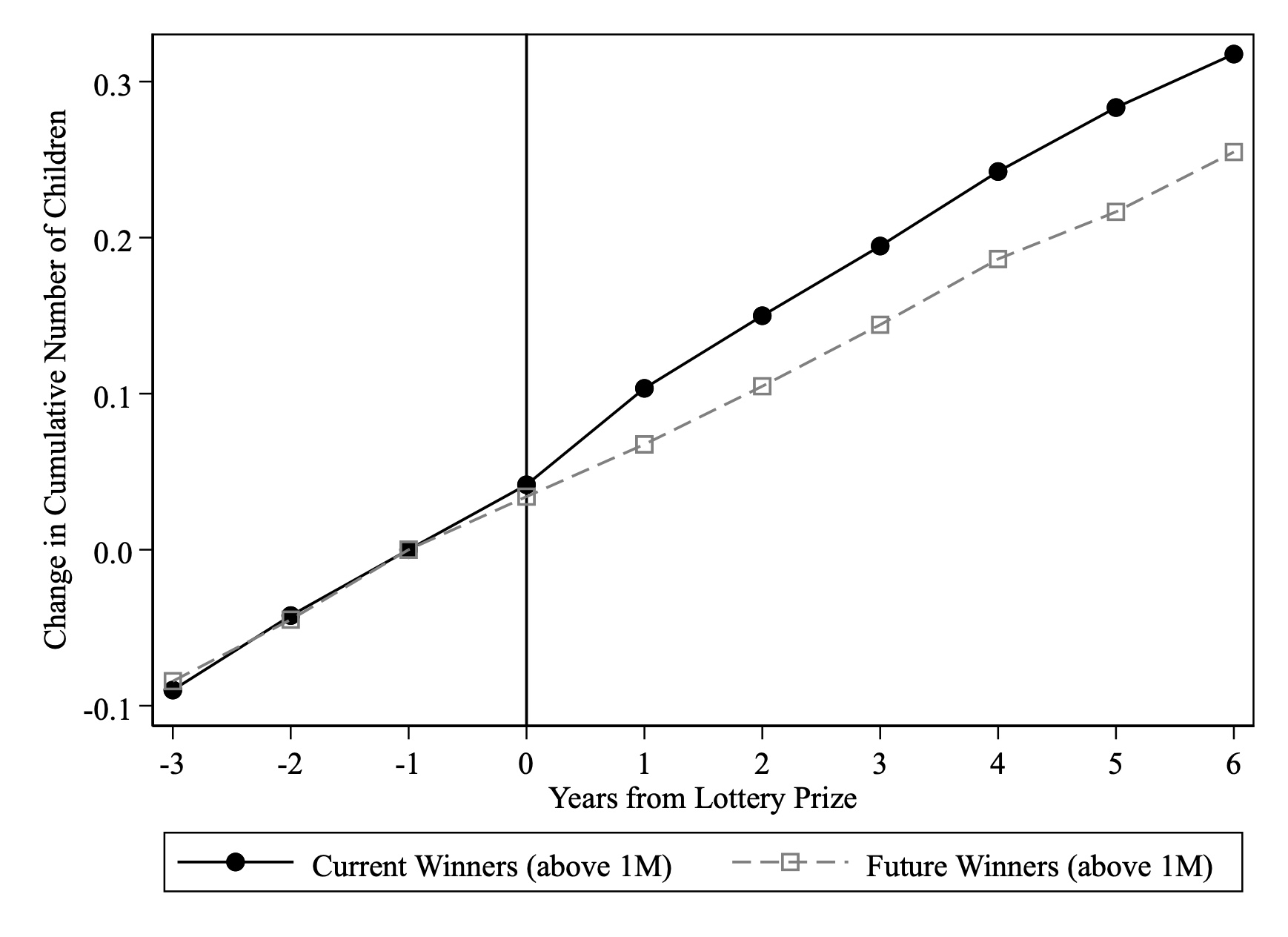}
	\fontsize{10}{10pt}\selectfont
	\flushleft
	\emph{Notes:} This figure compares the trend in the number of cumulative children from three years before to six years after the time of winning a lottery prize. The solid line with circular symbols stands for current winners who won above NT\$ 1M, and the dashed line with square symbols stands for future winners who won the same amount in prize money. The vertical axis displays the outcomes (the number of cumulative children) relative to the baseline year (one year previous to the (placebo) lottery-winning year) for each group. The horizontal axis refers to the number of years from the (placebo) lottery-winning year.
\end{figure}

\begin{figure}[H]
	\centering
	\caption{Effect of a Ten Million NT\$ Lottery Prize on Fertility} \label{fig.DDD}	 
		\includegraphics[width=0.8\textwidth]{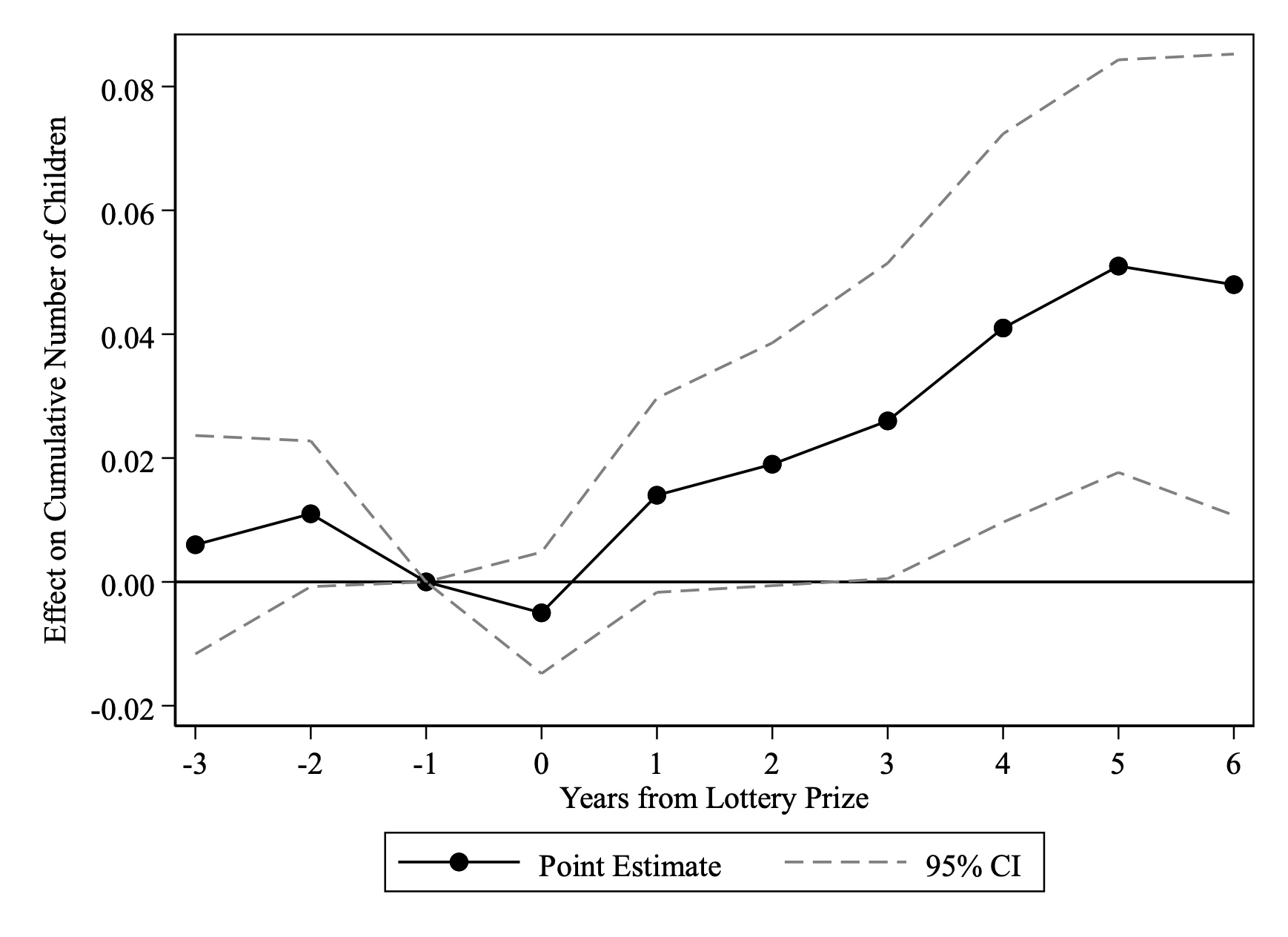}
	\fontsize{10}{10pt}\selectfont
	\flushleft
	\emph{Notes:} This figure displays the estimated coefficients of $Current_{i} \times Prize_{i} \times \mathbf{I}[t=L_{i}+s]$ from Equation (\ref{child_event}). The outcome of interest is the cumulative number of children. The solid line denotes the point estimates. The dashed line denotes the 95\% confidence interval. The horizontal axis refers to the number of years from the (placebo) lottery-winning year.
\end{figure}

	In this section, we discuss the main results. Figure \ref{fig.DDD} shows the estimated $\gamma_{s}$ of our DDD regression (Equation (\ref{child_event})), i.e., the effect of a 10 million NT\$ windfall on cumulative fertility. First, we find that the estimated coefficients in the pre-winning period ($s=-3, -2$) are very small and statistically insignificant, thereby suggesting that pre-trends run parallel. Consistent with the graphical evidence in Figure \ref{fig.trend}, the estimated $\gamma_{s}$ indicates that the receipt of a large cash windfall can stimulate fertility immediately, and the effects persist for at least 6 years. 
	
	As our primary focus lies on the total number of children, we use the estimate from the sixth year post-lottery win ($s=6$) to encapsulate the effect of lottery wins on fertility. Table \ref{tab.main} documents the DDD estimates, respectively. We commence by introducing the estimate from a basic model without any controls (Column (1)). We then progressively introduce fixed effects for the winner's age, year-fixed effects, individual characteristics prior to the win, outcome variables prior to the win, and past lottery winning amounts (Columns (2) to (6)). The stability of the estimates across various specifications is reassuring and provides robustness to our results.

	Our preferred specification is Column (6) in Table \ref{tab.main}, which includes all covariates. It suggests that winning a prize of 10 million NT\$ leads to a significant increase in the number of births by 0.05. In other words, for every 100 winners who won 10 million NT\$, five more children were born by the sixth year following the win than what would have been born in the absence of receiving a major prize. This corresponds to a 15\% increase in the baseline change in the number of children born between $s=-1$ and $s=6$ for the comparison group (i.e., future winners) at 0.32. 

To assess the sensitivity of fertility behaviors to a change in wealth, we further calculate the corresponding elasticity of fertility with respect to wealth. The change in household wealth between $s=-1$ and $s=6$ for future winners (i.e., comparison group) is around 3.88 million NT\$. We utilize this figure as the potential amount of wealth that treated individuals could have accumulated, given they did not receive a cash windfall. Consequently, this suggests that winning a 10 million NT\$ windfall translates into a 258\% increase in potential wealth accumulation for current winners (i.e., the treatment group). Given the above information, our result indicates that the total wealth elasticity of fertility is around 0.06.

	In sum, our results demonstrate a positive effect of cash windfalls on fertility. The positive income/wealth effect is consistent with the central proposition made by the neoclassical economic theory of fertility, in that children are normal goods, as proposed by Gary \citet{becker1960economic,becker1965theory}. 
		
\subsection{Falsification Tests and Robustness Checks}
	
\begin{figure}[H]
	\centering
	\caption{Permutation Test: Randomly Assigned Lottery Prize}	\label{fig.placebo}	 
	\begin{subfigure}[b]{0.6\textwidth}
		\caption{Main Results and Placebo Estimates}\label{fig.placebo.event}
		\includegraphics[width=1\textwidth]{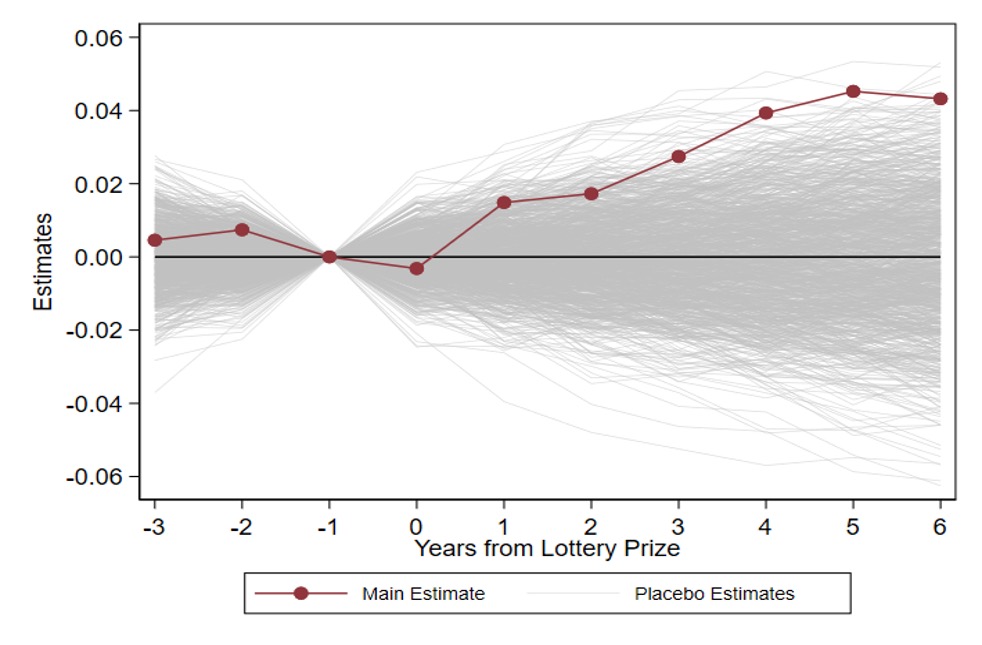}	
	\end{subfigure}
	\\
	\begin{subfigure}[b]{0.6\textwidth}
		\caption{The Estimated Coefficient of $Prize_{i} \times \mathbf{I}[t=6]$ and Placebo Estimates}\label{fig.placebo.hist}
		\includegraphics[width=1\textwidth]{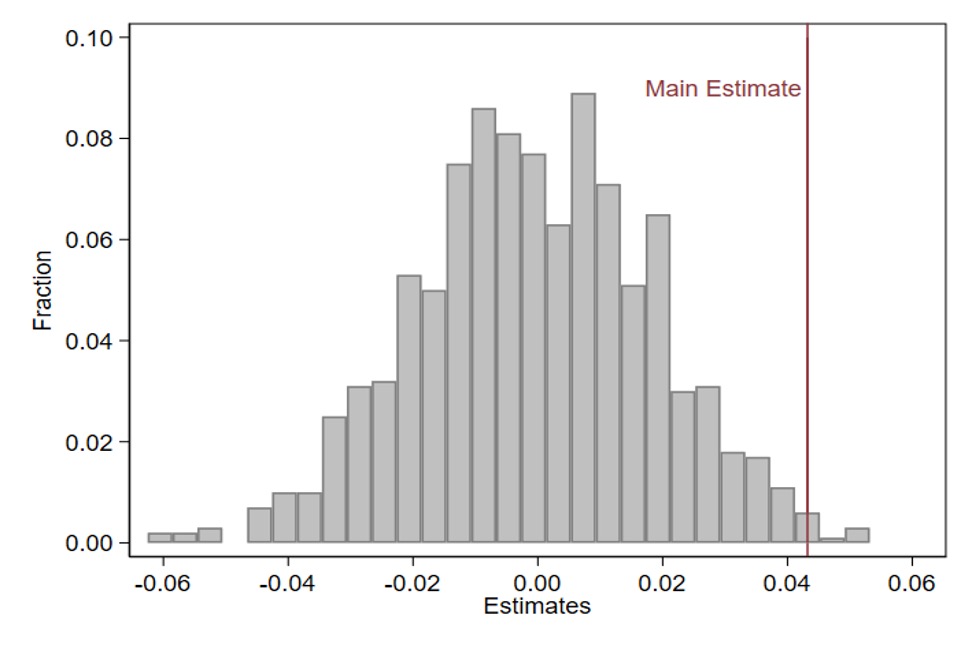}
	\end{subfigure}
	\\
	\fontsize{10}{10pt}\selectfont
	\flushleft
	\emph{Notes:} This figure shows the distribution of 1,000 placebo estimates. Specifically, we randomly permute lottery prizes and attach them to each winner. Then, we use these ``pseudo'' prizes to estimate Equation (\ref{child_event}). We repeat the above procedures 1,000 times to obtain the distribution of pseudo estimates. Figure \ref{fig.placebo.event} compares the real estimate (bold line with a circular symbol) with these fake ones (thin lines, gray in color). As we mainly focus on the estimated coefficient of $Current_{i} \times Prize_{i} \times \mathbf{I}[L_{i}+6]$, which summarizes the effect of the 10 Million NT\$ lottery win on total fertility, Figure \ref{fig.placebo.hist} shows the real estimates (vertical line) and distribution of pseudo ones (histogram) for $\gamma_{6}$.
\end{figure}

	In this section, we first implement a series of falsification tests for our preferred specification (i.e., Column (6) in Table \ref{tab.main}). Specifically, we randomly permute lottery prizes and attach them to each winner. Then, we use these ``pseudo'' prizes to define variable $Prize$ and estimate Equation (\ref{child_event}). We repeat the above procedures 1,000 times to obtain the distribution of pseudo estimates. Figure \ref{fig.placebo.event} compares the real estimate (bold line with circle symbol) with these fake ones (thin lines, gray in color). The result suggests that the real estimates of $\gamma_{s}$ are much larger than the pseudo ones in the post-winning period. As we mainly focus on the estimated coefficient of $Current_{i} \times Prize_{i} \times \mathbf{I}[t=L_{i}+6]$, which summarizes the effect of a lottery win on total fertility, Figure \ref{fig.placebo.hist} illustrates real estimates (vertical line) and the distribution of pseudo ones (histogram) for $\gamma_{6}$. The result suggests that the real $\gamma_{6}$ estimate is exceptionally larger than any fake ones. Specifically, the permutation p-value is 0.003. In sum, the placebo test confirms that significant estimates in our main results are unlikely to be chance findings.
	
	Next, we carry out a range of robustness checks for our main results, the outcomes of which we display in Table \ref{tab.robust}. Again, we use estimated effects at the sixth year post-lottery win ($s=6$) to summarize the wealth effect on total fertility so that Table \ref{tab.robust} only displays the estimated $\gamma_{6}$. Our main results are based on all lottery wins between 5,000 NT\$ to 100 million NT\$. In conducting the robustness checks, we confine our analysis to winnings amounting to 30,000 NT\$ or more (see Column (1)). This step is taken to verify that the main results are not driven by the large volume of individuals winning small prizes. We also lower the minimum win threshold to 2,000 NT\$ (see Column (2)), which is the smallest win we can observe. Despite this alteration, our results prove to be robust.
	
	In the main specification, we restrict the maximum lottery win amount to 100 million NT\$. To check the robustness of our results, we examine two alternative maximum win amounts: 80 million NT\$ (see Columns (3)) and 120 million NT\$ (see Column (4)). The estimate when using smaller maximum win amounts is greater in magnitude compared to the main result, while the estimate when using larger maximum win amounts is similar to our main results. Still, all estimates remain statistically significant. 
	
	Another concern for the main estimate is that our sample only consists of lottery winners, as the characteristics of these people could be different from those of the general population. Table \ref{tab.descriptive} indicates that the lottery sample was slightly older. Consistent with this fact, they are more likely to be married and employed than the general population. In order to investigate this issue, we first re-weight the sample to make these characteristics similar to those of the general population in Taiwan.\footnote{We use the post-stratification weighting technique and match the marital status, age, earnings, and asset stratifications for our lottery sample and the population, the latter of which is defined as individuals aged 20 to 44 from 2007 to 2012 (same as our winning years) in Taiwan. This leads to 11 million observations, which is around half of the nation's population. We randomly assign a placebo-winning year to the population and use their characteristics as one year prior to the placebo-winning years.} After re-weighting, although differences in observable characteristics between the lottery sample and the population are still statistically significant due to the large sample size, the magnitudes become much smaller (see Table \ref{tab.descriptive.weighted} in the Online Appendix \ref{app: add_t_f}). The sizes of these differences, as the proportion of the population means, are mostly below 10\%. Column (5) of Table \ref{tab.robust} suggests that the result based on population re-weighting is statistically significant but gives a slightly smaller estimate (i.e., 0.042). 
	
	In our main specification, we control for fixed effects based on the age of the winner. However, female age is more likely to be the key factor determining fertility behavior, so in this robustness check, we include fixed effects (separately) for male and female ages.\footnote{For male winners, the female age is defined as their pre-lottery spouse's (if present) age; for female winners, the male age is defined as their pre-lottery spouse's (if present) age; for female winners; for winners with no pre-lottery spouse, we create a dummy indicating the missing age for their spouse. Taiwan legalized same-sex marriage in 2019. The last observed year in our analysis is 2018, so there is no case when a winner and their spouse is of the same sex.} As indicated in Column (6) of Table \ref{tab.robust}, the estimate (0.050) does not deviate from our main results. 
	
	Additionally, we implement a robustness check, incorporating individual fixed effects to account for any unobserved, time-invariant disparities between individuals that may influence fertility decisions, such as the preference to have children and the ability to get pregnant. The resulting estimation (0.045), found in Column (7) of Table \ref{tab.robust}, aligns closely with our primary findings.
	
	In the main specification, we define the winning year as the year of the first lottery win, thus allowing individuals to win multiple times. Since subsequent lottery wins could also influence fertility, we check the robustness of only including individuals who won the lottery once during the sample period. Column (8) of Table \ref{tab.robust} reports estimates using the restricted sample of single-time winners. The estimate is 0.050, which is close to our main result. 
 
	Our data consists of six cohorts with different treatment timings, namely, treated individuals winning lottery prizes in a given year from 2007 to 2012. Several recent studies \citep{de2020two,callaway2021difference,goodman2021difference,baker2022much,sun2021estimating} suggest that if treatment effects are heterogeneous across treated cohorts, conventional DID estimates could be biased. To address this concern, we individually estimate the lottery effect for each cohort and then average these estimates. In particular, we compare winners who secured the lottery prize in 2007 with the corresponding future winner cohort (those who won the lottery prize in 2014, with the placebo winning year set at 2007). We carry out a similar estimation separately for each cohort, and subsequently, we calculate the average estimate, weighted by the sample size of each cohort. We estimate the standard error through 1,000 times bootstrapping (re-sampling with a replacement within lottery cohorts). This approach ensures that we avoid comparing observations from different treatment timings and mitigates the bias that could arise from a staggered DID design. Column (9) of Table 3 presents the estimated effect as 0.046, which closely aligns with our main result. Overall, the evidence in this section demonstrates that our primary estimate remains robust across various sample selection criteria and empirical specifications.
		
\subsection{Shift in Fertility Timing or a Change in Total Fertility}
	
\begin{figure}[H]
	\centering
	\caption{Effect of a Ten Million NT\$ Lottery Prize on Fertility: Young and Middle-aged Winners}\label{fig.age}
	
	\begin{subfigure}[t]{0.6\textwidth}
		\caption{Young Winners}\label{fig.young}
		\includegraphics[width=1\linewidth]{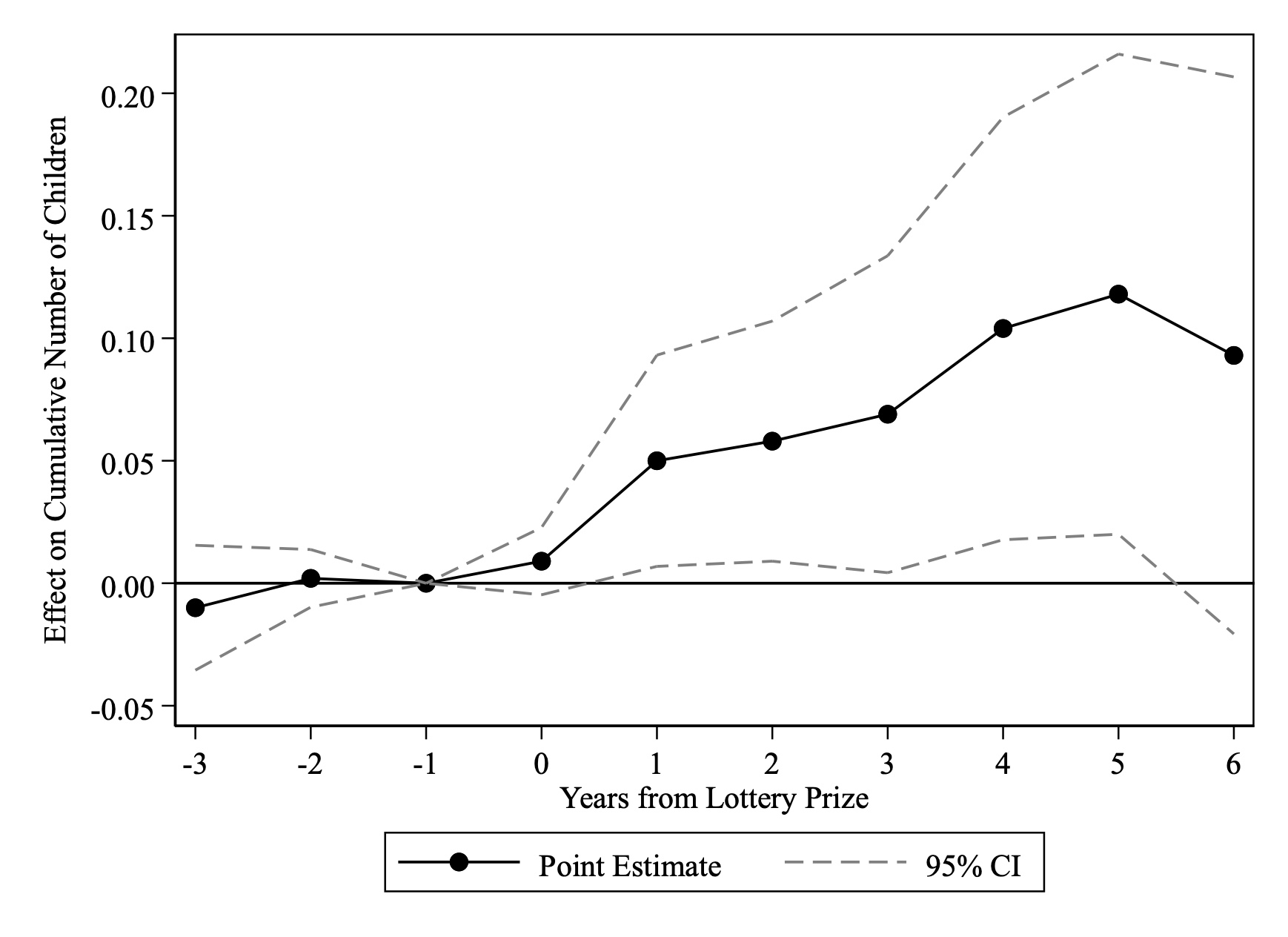}
	\end{subfigure}
	\\
	\begin{subfigure}[t]{0.6\textwidth}
		\caption{Middle-aged Winners}\label{fig.middleage}
		\includegraphics[width=1\linewidth]{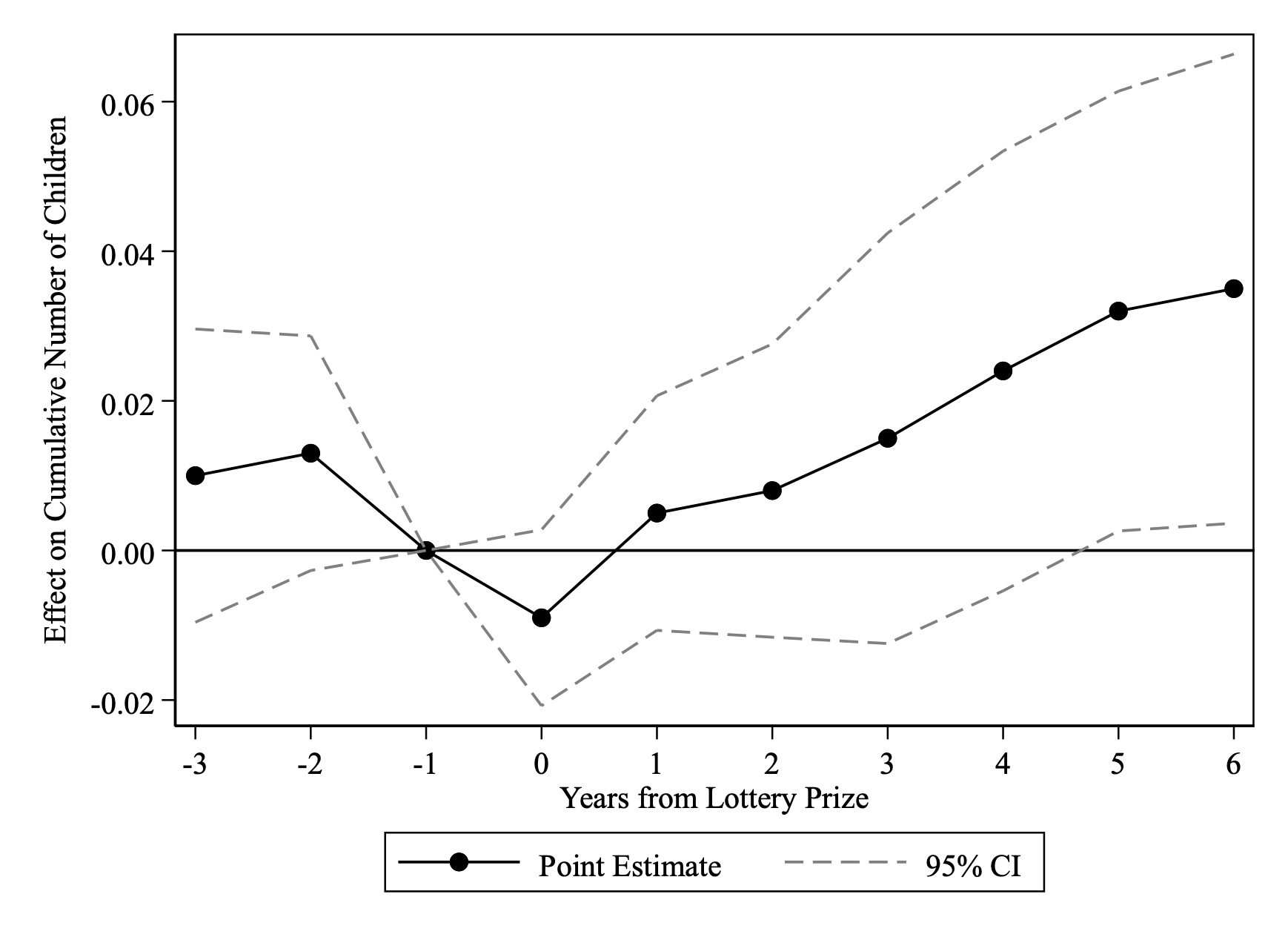}
	\end{subfigure}
	\fontsize{10}{10pt}\selectfont
	\flushleft
	\emph{Notes:} These two figures display the estimated coefficients of $Current_{i} \times Prize_{i} \times \mathbf{I}[t=L_{i}+s]$ from Equation (\ref{child_event}). The outcome of interest is the cumulative number of children. The solid line denotes the point estimates. The dashed line denotes the 95\% confidence interval. The horizontal axis refers to the number of years from the (placebo) lottery-winning year. Figure \ref{fig.young} shows the results for young winners (i.e., the age of the winner is below 30 years old). Figure \ref{fig.middleage} shows the results for middle-aged winners (i.e., the age of the winner is above 30 years old).
\end{figure}

	Our main results could reflect either shifts in fertility timing or changes in total lifetime fertility. To explore this distinction, we examine heterogeneous responses by age group in Figure \ref{fig.age}. Young winners are defined as those aged 20-29 when they won the lottery prize (see Figure \ref{fig.young}). Middle-aged winners are those aged 30-44 at the time of winning (see Figure \ref{fig.middleage}).
		
Figures \ref{fig.young} and \ref{fig.middleage} present the estimated effect of winning a 10 million NT\$ prize on fertility for young and middle-aged lottery winners. Both age groups exhibit significant increases in fertility, with young winners having 0.093 more children and middle-aged winners having 0.035 more children by the sixth year after winning the lottery. The larger effect for young winners aligns with expectations given declining fertility rates with age. Notably, the average age of middle-aged winners is 36 prior to winning the lottery and reaches 43, an age of low fertility probability, by the end of the sample period. Nonetheless, fertility continues to increase through the sixth year for middle-aged winners, implying that their lifetime fertility likely rises due to the lottery windfall.

Moreover, in Figure \ref{fig.b3a8} of Online Appendix \ref{app: add_t_f}, we replicate the core analysis using only four lottery-winning cohorts. This modification enables us to follow winners for up to eight years after their winning year. We find that the effect of lottery wins on fertility remains positive and persistent. Although the reduced sample size leads to larger standard errors, the estimates are still marginally statistically significant ($p<0.10$). This suggests that the observed changes in fertility are more likely due to overall shifts in total fertility rather than temporary changes in fertility timing.
	
\subsection{Heterogeneous Effects: Financial Resources}\label{sec:subgroup}

	In this section, we examine whether insufficient financial resources pose a barrier to having children. Our main results suggest that children are normal goods - that is, individuals derive consumption value from having children and are more likely to do so when they have more resources. This leads to the prediction that the positive effect of a cash windfall on fertility should increase in line with the size of a cash windfall. 
	
	To examine this prediction, we implement a design that classifies wins by prize amount, including extremely large prizes over 100 million that were excluded from earlier analyses. This allows us to explore the threshold level of resources needed to impact fertility decisions, as well as to measure the upper bound effects of very large windfalls. Specifically, we estimate the following regression:
	
	\begin{align}\label{child_event_step}
	B_{it}&= \alpha_{0} Large_{i}  + \sum_{s\neq-1} \kappa_{s} \cdot \mathbf{I}[t=L_{i}+s]    +\sum_{s\neq-1}  \lambda_{s} \cdot Large_{i} \times \mathbf{I}[t=L_{i}+s] \nonumber   \\ 
	&+ (\alpha_{1} + \alpha_{2} Large_{i}  + \sum_{s\neq-1} \beta_{s} \cdot \mathbf{I}[t=L_{i}+s]    +\sum_{s\neq-1}  \gamma_{s} \cdot Large_{i} \times \mathbf{I}[t=L_{i}+s] ) \times Current_{i} \nonumber   \\
	& + a_{it} + \theta_{t}  + \boldsymbol{X_{i}} \psi  + \varepsilon_{it} 
	\end{align}
	
	The notation in Equation (\ref{child_event_step}) is similar to Equation (\ref{child_event}) but replaces the continuous lottery amount variable $Prize$ with a binary indicator $Large$ for large versus small wins. We define small winners as those receiving between 5 and 10 thousand NT\$, and we categorize large winners into five prize groups: 1) 10 to 50 thousand NT\$; 2) 50 to 500 thousand NT\$; 3) 500 thousand to 5 million NT\$; 4) 5 to 100 million NT\$ and 5) 100 million NT\$ or more. 
	
	The indicator $Large$ takes a value of 1 for individuals winning prizes in one of the above five large prize groups, and 0 for those winning small prizes under 10 thousand NT\$. On average, the small prize group won around 8 thousand NT\$. The key coefficients of interest, $\gamma_s$, capture the effects of winning a larger lottery prize compared to small wins under 10 thousand NT\$, which is the omitted base category. By segmenting large prizes into categorical groups, we can estimate differential impacts based on the magnitude of the windfall.
	
	Table \ref{tab.prizegrp} reports heterogeneity in fertility responses to windfall gains according to the size of a lottery win. For moderate resource shocks between 10 thousand and 5 million NT\$, the impact on fertility is small, with coefficient estimates around 0.01--0.03 (Columns (1) to (3)). However, effects grow substantially once prize amounts reach 5--100 million NT\$, with estimates increasing to approximately 0.1 (Column (4)). The response continues to rise for windfalls and peaks at the highest level of 0.348 for Jackpot wins exceeding 100 million NT\$ (Column (5)).
	
	Consistent with the above results, we also find stronger sensitivity to cash windfalls for less wealthy individuals. Table \ref{tab.financial} examines heterogeneous effects relative to individual financial resources. The results in Columns (1) and (2) indicate that individuals with no deposit appear to be driving the positive fertility effect of cash windfalls. For winners with no deposit, receipt of a 10 million NT\$ lottery prize significantly raises the cumulative number of children by 0.066 in the sixth year after winning (Column (1)). In contrast, the estimate in Column (2) suggests that fertility responses are small and statistically insignificant for those with cash at hand. Similar results are found when we define financial resources by liquid assets (deposit plus stock). Columns (3) and (4) reveal that the positive fertility effect of windfalls is primarily driven by individuals with zero liquid assets prior to winning. Those having no liquid assets experienced a 0.1 increase in the number of children in the sixth year after winning (Column (3)). In contrast, those with some liquid assets experienced a null effect from receiving a lottery prize. Columns (5) and (6) derive the same results when using total assets (liquid assets plus real estate less house loan debt). Overall, the evidence reveals a pattern of larger fertility responses for individuals receiving larger cash windfalls and those with fewer pre-existing financial resources. These findings imply that individuals make the choice to have children once they have accumulated sufficient wealth to finance the cost of raising them.

\subsection{Heterogeneous Effects: Extensive vs. Intensive Margins}
	
	To examine whether cash windfalls influence fertility through either the extensive margin (having children or not) or the intensive margin (having additional children), we analyze parenthood status in Table \ref{tab.household} Columns (1) and (2), which compare individuals with and without children the year before winning ($s=-1$). The results show that the main effect is driven by the extensive margin, in that childless individuals receiving a 10 million NT\$ windfall have 0.093 more children by the 6th year after winning. However, for individuals with children already, the windfall does not change fertility behavior.
	
	Parenthood and marital status are closely linked, as childless individuals are typically unmarried. Thus, Columns (3) and (4) explore response differences by marital status. Aligning with the parenthood findings, the fertility increase mainly comes from single individuals (Column (3)). The effect is insignificant for married couples (Column 4). In the last two columns of Table \ref{tab.household}, we combine parenthood and marital status and extend the heterogeneity analysis according to whether couples have pre-win children, or not. Our results indicate that the fertility behaviors of childless couples are very responsive to lottery wins. Receiving a 10 million NT\$ windfall significantly increases the number of children ever had by childless couples by 0.357 at the end of the sixth year following a win, albeit for couples with children already, the prize money does not influence fertility.
	
	In summary, we find that a cash windfall raises fertility levels, mostly along the extensive margin. However, the receipt of a large cash amount triggers only childless individuals to give birth and has little impact on subsequent fertility for those who already have children. This implies that parents likely prioritize investing in their existing children rather than having more. Our finding aligns with Becker's conjecture that the quantity of income elasticity should be small when parents consider the quantity-quality trade-off \citep{becker1960economic, becker1965theory,becker1973interaction, doepke2015gary,li2008quantity}. To test this hypothesis, in the next section, we examine whether lottery wins improve child quality, measured by college attendance.

\section{The Effect of Cash Windfalls on Other Related Outcomes}\label{sec:other_o}	
	
\subsection{Children's College Attendance}

	Our results so far have demonstrated that lottery windfalls do not increase subsequent fertility for parents who already have children before winning. To understand this negligible impact, we perform a complementary analysis examining child quality as the outcome. 

	Becker's theory of fertility suggests that parents face a trade-off between quantity and quality when making decisions about having children \citep{becker1960economic, becker1965theory,becker1973interaction, doepke2015gary,li2008quantity}. When additional resources become available, they may opt to invest that money into improving the quality of their existing children, rather than having more. To test this hypothesis, we utilize children's college attendance as a proxy measure for child quality. Following \citet{bulman2021parental}, we compare two groups: children whose parents won lottery prizes before the child finished high school (age 19), and those whose parents won after age 19 (i.e., children of future winners). The latter group serves as a control to absorb unobserved differences 
	between households receiving larger and smaller windfalls.\footnote{Because college attendance is typically a one-time event in a child's late teens, we cannot observe the same child's attendance both before and after their parent won the lottery. Thus, we are not able to use the DDD design as we did in the main analysis.}
	
	Therefore, the estimation sample includes only individuals with children turned 19 before or after a lottery win. The final sample size is 80,661 children within 58,432 winners. Utilizing this sample, we examine whether lottery windfalls increase subsequent college attendance for winners' pre-existing children. If windfalls improve college attendance, it would suggest that parents invest money into child quality when they opt not to have more children. Specifically, we compare differences in college attendance rates between children of current and future winners who won larger or smaller prizes by estimating the following regression:
	
	\begin{align}\label{child_q}
		E_{ij}&= \delta_{1} Current_{j} + \delta_{2} Prize_{j} + \rho \cdot Current_{j} \times Prize_{j}  + \gamma_{c} + \theta_{t} + \boldsymbol{X_{j}} \psi + \boldsymbol{Z_{i}} \nu  + \varepsilon_{ijt} 
	\end{align}

\noindent where $E_{ij}$ represents the outcome of interest for child $i$ whose parent is a winner $j$---a series of dummy variables indicating whether child $i$ has ever attended college and the type of college attended, measured at age 19. We examine three outcomes: 1) ever attending any college; 2) ever attending a domestic college; and 3) ever attending an overseas college. 

	$Current_{j}$ is a dummy indicating whether parent $j$ is a current winner ($Current_{j}=1$) or a future winner ($Current_{j}=0$), meaning they won prizes before or after their child turned 19 years old. The variable $Prize_j$ is a continuous measure of the amount won by parent $j$. The coefficient $\delta_{1}$ on $Current_{j}$ captures any fixed differences in college outcomes between children whose parents won prizes before versus after age 19. The coefficient $\delta_{2}$ on $Prize_{j}$ controls for potential heterogeneity arising from parents winning different prize amounts. The coefficient of interest $\rho$ on the interaction between $Current_{j}$ and $Prize_{j}$ represents the effect of lottery wins on college attendance.

	To isolate the impact of lottery prizes, the model includes fixed effects $\gamma_{c}$ for the child's birth year, to absorb cohort differences. Calendar year fixed effects $\theta_t$ for when the child turns 19 are also included to account for contemporaneous factors affecting overall college attendance. We further control for winner (parent) characteristics $X_{j}$\footnote{Characteristics are the same as the covariates included in Column (6) of Table \ref{tab.main}, i.e., our main analysis of lottery impact on fertility} and child characteristics $Z_{i}$\footnote{Characteristics include a child's gender, birthplace, birth order, and birth month.} to address outcome heterogeneity arising from these observable factors.

	Table \ref{tab.education} presents the estimated effects of lottery wins on college attendance for the children of winners. Columns (1) to (3) suggest no significant impact of lottery wins on the likelihood of college attendance. In Taiwan, like other countries, some students may choose to study overseas (e.g., in the US) for higher-quality education, where tuition and living costs are much higher than domestic options. Receiving a cash windfall could enable parents to afford to send their children abroad to study. To examine this notion, we disaggregated college attendance into domestic (Columns (4) and (6)) and overseas (Columns (7) and (9)). The results indicate lottery wins have no significant impact on attending domestic college.\footnote{This insignificant impact could be explained by the relatively low cost of college attendance in Taiwan, i.e., on average, 58 thousand NT\$ (around 2 thousand US\$) for public universities and 110 thousand NT\$ (around 3.6 thousand US\$) for private universities per academic year.} However, we find a 10 million NT\$ windfall significantly increases the probability of studying abroad for undergrads by 1.7 percentage points. This is a sizeable change, equivalent to a 121\% increase relative to the baseline mean for studying abroad.

In summary, while lottery prizes do not appear to influence overall college attendance, the winnings do significantly increase the likelihood of children studying overseas. This suggests that cash windfalls enable winners to afford the higher costs of international education for their children, which is perceived as higher quality.

\subsection{Marriage Decisions}
	
	Our results so far reveal that cash windfalls increase fertility, primarily through the extensive margin. Fertility and marriage decisions are often interrelated \citep{upchurch2002nonmarital,baizan2003cohabitation,aassve2006employment,marchetta2016role}, especially in East Asian societies where people typically marry before having children \citep{Myong_etal_2021}. One potential mechanism for the fertility effect is that windfalls induce marriage. We examine whether lottery wins influence marriage decisions by replacing the outcome in Equation (\ref{child_event}) with a dummy indicating ``ever married.''\footnote{In this analysis, we use ``ever married'' as the dependent variable to capture the effect of a lottery win on the decision to marry, rather than the timing of marriage. Nevertheless, the results are robust when changing the outcomes to ``being married'' at the observed year.}
	
\begin{figure}[H]
	\centering
	\caption{Effect of a Ten Million NT\$ Lottery Prize on Marriage} \label{fig.married}	 
	\includegraphics[width=0.8\textwidth]{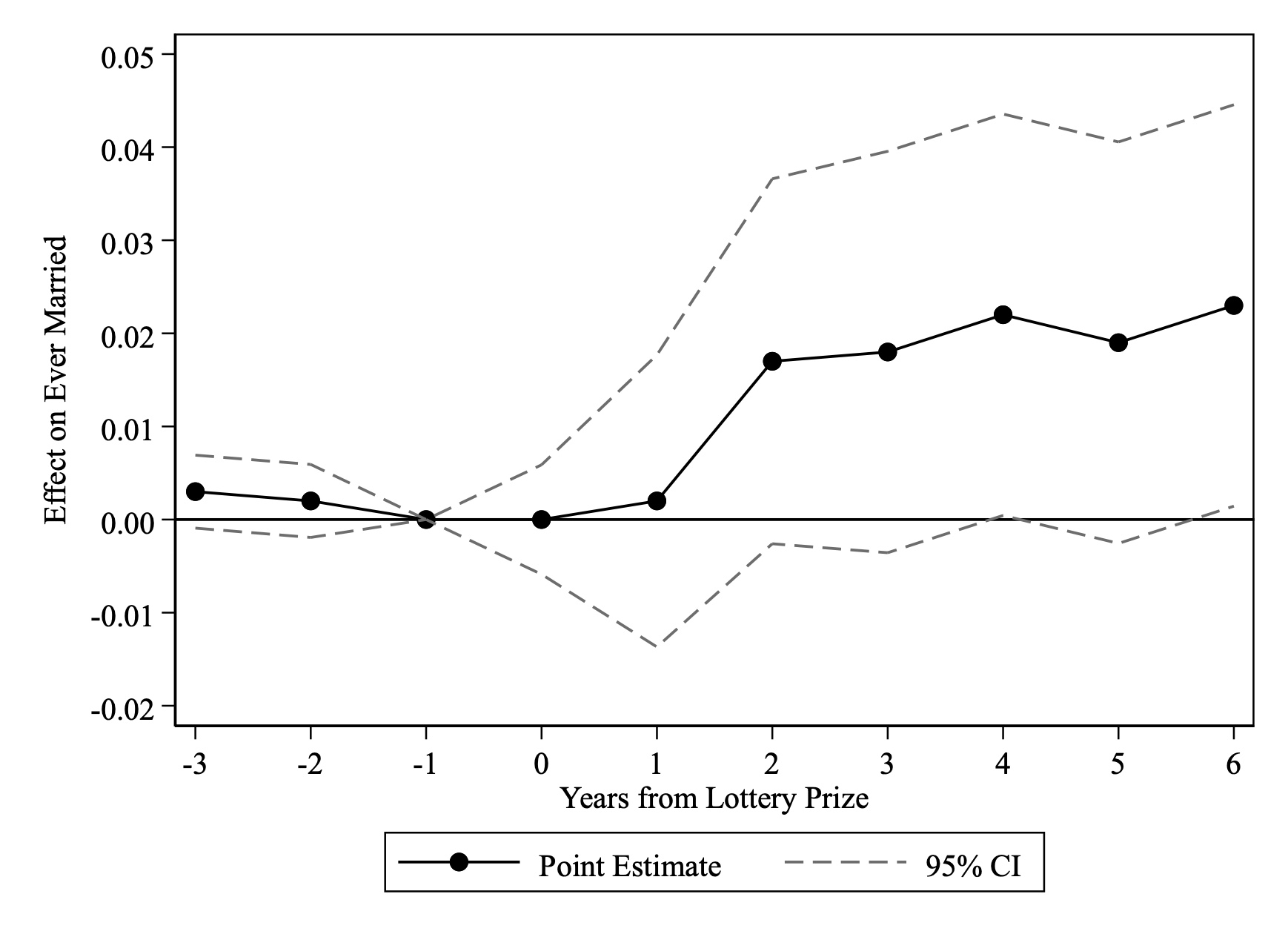}
	\fontsize{10}{10pt}\selectfont
	\flushleft
	\emph{Notes:} This figure displays the estimated coefficients of $Current_{i} \times Prize_{i} \times \mathbf{I}[t=L_{i}+s]$ from Equation (\ref{child_event}). The outcome of interest is ever getting married. The solid line denotes the point estimates. The dashed line denotes the 95\% confidence interval. The horizontal axis refers to the number of years from the (placebo) lottery-winning year.
\end{figure}

	Figure \ref{fig.married} displays the dynamic DDD estimates for the effect of windfalls on the decision to marry. We find that people are more likely to get married for the first time after receiving a substantial windfall, and this positive effect persists over time. Table \ref{tab.married} summarizes the results, illustrating stable estimates across specifications. Our preferred estimate implies that a 10 million NT\$ windfall significantly increases the probability of getting married within a six-year horizon by 2.3 percentage points---a sizeable 14\% increase compared to the baseline trend.\footnote{The baseline trend refers to changes in the share of future winners who ever got married between $s=-1$ and $s=6$, which is 16.1\%."}

To investigate how much of the observed effect on fertility can be attributed to changes in marriage behavior, we conducted a causal mediation analysis, following the approach of previous studies \citep{hsia2021causal,breivik2022career}. However, due to our reliance on a single source of exogenous variation (lottery wins), and the fact that both outcomes (marriage and fertility) were determined within the same period, we lacked the specific variation needed to clearly isolate the impact of marriage on fertility.  Therefore, the mediation analysis should be interpreted cautiously, albeit it provides a useful insight into whether the marriage mechanism can potentially explain treatment effects.

Specifically, we assumed lottery wins have both direct effects on fertility and indirect effects through influencing marriage behavior. Indirect effects can be obtained by decomposing the effect of lottery wins on fertility $\gamma_{s}$ in Equation (\ref{child_event}) into three components: 1) the effect of marriage on fertility; 2) the effect of lottery wins on marriage; and 3) the unexplained part of the lottery effect (i.e., direct effect). The product of the first two components can be viewed as an increase in fertility caused by lottery wins through changing marriage behavior. Following \citet{hsia2021causal} and \citet{breivik2022career}, we first estimate the impact of marriage on fertility while controlling for the effect of lottery wins by adding the marriage mediator variable $M$ to Equation (\ref{child_event}):
	
	\begin{align}\label{child_event_med}
		B_{it}&= \pi M_{i} + \alpha_{0} Prize_{i}  + \sum_{s\neq-1} \kappa_{s} \cdot \mathbf{I}[t=L_{i}+s]    +\sum_{s\neq-1}  \lambda_{s} \cdot Prize_{i} \times \mathbf{I}[t=L_{i}+s] \nonumber   \\ 
		&+ (\alpha_{1} + \alpha_{2} Prize_{i}  + \sum_{s\neq-1} \beta_{s} \cdot \mathbf{I}[t=L_{i}+s]    +\sum_{s\neq-1}  \gamma_{s} \cdot Prize_{i} \times \mathbf{I}[t=L_{i}+s] ) \times Current_{i} \nonumber   \\
		& + a_{it} + \theta_{t}  + \boldsymbol{X_{i}} \psi  + \varepsilon_{it} 
	\end{align}
where $M$ indicates whether an individual has ever entered into marriage. Our estimation shows that the effect of being married on fertility was approximately 0.51. Next, we multiply this estimate by the estimated effect of lottery wins on marriage, as shown in Table \ref{tab.married} (0.023). This calculation indicates that the indirect effect of lottery wins on fertility through its influence on the marriage rate is approximately 0.012, thereby accounting for 25\% of the total impact (0.048).\footnote{The effect of marriage on fertility is 0.51, and the effect of a lottery win on marriage is 0.023. The product of these two effects is $0.51 \times 0.023 = 0.012$.}
		
\section{Discussion} \label{sec:discussion}
	
	In this section, we discuss the relationship between our results and the findings from previous studies. Furthermore, we discuss the policy implications of our findings.
	
\subsection{Comparison with Early Studies Using a Quasi-Experimental Design}

	The early literature examines the effect of financial resources on fertility, using various sources of income/wealth shocks for identification \citep{ager2019structural, alam2018income,huttunen2016effect,lovenheim2013family,black2013children,lindo2010children}. Estimated effects differ dramatically across these early studies. The estimated income/wealth elasticities of fertility range from $-0.07$ to $0.65$. For example, \citet{lindo2010children} finds that a negative income shock generated by a husband losing his job significantly reduces total fertility, while \citet{huttunen2016effect} indicates that male job loss, which results in a larger decrease in family income than female job loss, has a negligible impact in this regard. An important concern when estimating the causal effects of income/wealth shocks on fertility decisions is that unobserved factors may affect the likelihood of both shocks and outcomes. In addition, due to variations in the sources used for identification, previous studies usually restrict their sample to married couples, since the analysis is based on variations in a husband's labor income \citep{alam2018income,huttunen2016effect,black2013children,lindo2010children}. Finally, estimating the heterogeneous effects of income/wealth on fertility decisions is challenging in these studies, since resource shocks used for identification are usually tied to specific subgroups such as homeowners \citep{lovenheim2013family}, job-losers \citep{huttunen2016effect,lindo2010children}, or workers in the mining industry \citep{black2013children}. Analysis of the wealth shocks from housing equity or job loss is restricted to individuals with a certain level of financial resources.\footnote{For example, \citet{lovenheim2013family} finds no fertility response for low-income individuals, which is somewhat puzzling. Given that all of the people in the sample own a house, it is hard to know whether their results indicate that either low-income individuals have lower wealth elasticity of fertility or that they respond differently to an increase in housing wealth than high-income people.}
	
	We complement these studies by utilizing relatively pure wealth shocks generated by lottery wins. The nature of shock generated by a lottery win enables us to include unmarried people, since everyone can participate in a lottery game. Our results reveal that encouraging single individuals to get married is an important mechanism through which income can influence fertility. Our research design relies on exogenous cash windfalls from lottery wins, which are not tied to individual baseline wealth or income. We find that the effect of wealth on fertility is greater for individuals with fewer financial resources.	
	
\subsection{Comparison to Lottery-based Studies}

	In this section, we compare our estimates with the results in two contemporaneous lottery-based studies from the US \citep{bulman2022effect} and Sweden \citep{cesarini2023fortunate}. To facilitate comparisons for these two studies, following \citet{cesarini2023fortunate}, we rescale our estimates to the effect of a windfall measured in units of 100,000 US\$ and focus on the fertility effect up to five years ($s=5$) after the lottery wins. 

Utilizing the estimates from Figure 2, our results indicate that lottery prizes of 100,000 US\$ increase the number of children by 0.015. The corresponding estimate in \citet{bulman2022effect} is 0.0007, suggesting that the effect of cash windfalls on fertility is very small. However, \citet{cesarini2023fortunate} establishes that a lottery win of 100,000 US\$ significantly increases cumulative fertility by 0.021 children within five years, which is slightly larger than our estimate. In addition, they provide a back-of-the-envelope calculation suggesting about 20--40 percent of male winners' fertility responses can be accounted for by their higher marriage rate. Although the approach is different, our causal mediation analysis also highlights the importance of marriage on fertility effects.

Comparing our results with the above lottery-based estimates reveals that the effect of wealth on fertility varies across countries, likely due to differences in policy and social contexts. This variation might be explained by three key factors. First, the pattern is associated with the level of public support for child-related expenses. Among OECD countries, Sweden has the highest spending (as a proportion of GDP) on family benefits, while the US ranks almost lowest (only exceeding Turkey). Specifically, Sweden spent 3.42\% of its GDP on family benefits in 2019, whereas the US spent only 0.62\% \citep{oecd2023family}.\footnote{Family benefits spending includes child-related cash transfers or in-kind support, public income support payments for parental leave, public spending on services for families, and tax benefits related to children.} Taiwan falls in the middle compared to Sweden and the US, spending around 0.89\% of its GDP on family benefits in 2019 \citep{DGBAS2020,FIA2022}. In addition, both the Swedish and Taiwanese governments highly subsidize education, even at higher education levels. In contrast, the US government provides limited tuition subsidies, with students bearing most of the costs through loans or out-of-pocket payments. On average, tuition and fees at public 4-year colleges in the US totaled 10,740 US\$ for in-state students in the 2019-2020 academic year.\footnote{Tuition fees for public universities in Taiwan are around 2,000 US\$ per year. In Sweden, public university education is tuition-free for Swedish citizens.}

Second, Taiwan has an extremely low fertility rate, hovering around 1, while Sweden and the US have higher and similar fertility rates at around 1.6--1.7. Furthermore, Taiwan's national income is relatively lower compared to Sweden and the US.\footnote{According to data from the International Monetary Fund, Taiwan had a GDP per capita of approximately 25,306 US\$ in 2019. This figure was lower than that of Sweden and the United States. Sweden, known for its high standard of living and strong social welfare system, had a GDP per capita of about 53,442 US\$. The United States, one of the world's largest economies, had an even higher GDP per capita, amounting to around 65,280 US\$.} Given our finding that the fertility effect is mainly driven by individuals with low liquid assets, these facts suggest that the significant impact of lottery wins on fertility in Taiwan could be partly attributed to the country's low baseline fertility rate and stage of development. A society with an already low fertility level and income may be more responsive to exogenous wealth gains.

Finally, social norms regarding marriage and childbearing may contribute to the variation across countries. In the U.S., lottery wins increase marriage but not fertility \citep{bulman2022effect}. In contrast, we find that getting married is an important mechanism driving the fertility effect in Taiwan, which reflects the social norm emphasis on childbearing within marriage commonly seen in Taiwan and many East Asian societies \citep{Myong_etal_2021}. The contrast across countries suggests that this cultural factor may shape how wealth gains influence fertility decisions, though more research is needed to understand these societal differences further.\footnote{Specifically, in East Asian countries, births outside of marriage are relatively rare compared to Western countries \citep{gietel2018changing,yeung2018families}, and delayed marriage greatly contributes to the low fertility rate \citep{jones2007delayed,straughan2008ultra}.}
 
\subsection{Policy Implications}

Our results, consistent with previous literature, indicate that wealth has a positive effect on fertility, and the estimated wealth elasticity of fertility is moderate.
These findings have several policy implications. First, the positive income/wealth effect implies that unconditional cash transfers or universal basic income can lead to moderate increases in fertility rates. Second, previous studies reveal that conditional child-related subsidies like baby bonuses and child allowances can influence fertility decisions \citep{milligan2005subsidizing,brewer2012does,cohen2013financial,gonzalez2013effect,kim2014lifetime,laroque2014identifying,ang2015effects,garganta2017effect,riphahn2017fertility,andersen2018can,malak2019baby,malkova2018can,stichnoth2020short,chuard2021baby, gonzalez2021cash,lyssiotou2021can,kim2022baby}. Theoretically, these subsidies affect fertility through two channels: 1) The income/wealth effect, i.e., by increasing family financial resources, subsidies make it more affordable to have children, and 2) The substitution effect, whereby subsidies lower the relative cost of having children versus other goods, which encourages fertility.

Given the modest income/wealth effect found in our study and other lottery-based designs, the effectiveness of child subsidies likely operates primarily through the substitution effect rather than the income/wealth effect. In other words, such policies are unlikely to dramatically increase fertility through purely increasing a family's financial resources. Rather, their impact stems from lowering opportunity costs of childbearing, which has important implications for policy design.

\section{Conclusion}\label{sec:conclusion}

	This study employs longitudinal administrative data on lottery winners in Taiwan to investigate the effect of cash windfalls on fertility behaviors. We find that a lottery win of 10 million NT\$ can significantly increase the number of children ever born by 0.05, which is equivalent to a 15\% increase from the baseline. The implied wealth elasticity of fertility is 0.06, which is consistent with the central proposition in Gary Becker’s neoclassical theory of fertility, in that children are normal goods, and so demand for the quantity of children should increase in line with individual financial resources \citep{becker1960economic,becker1965theory}.

	Additionally, less wealthy individuals exhibit greater fertility responses, suggesting that people have children after accumulating sufficient wealth. Cash windfalls primarily raise fertility by inducing first births among previously childless individuals (i.e., extensive margin) rather than making parents have more children (i.e., intensive margin). Lastly, our analysis reveals that lottery wins boost marriage, and approximately 25\% of the total fertility effect stems from increased marriage rates following lottery wins.	
	
	Our findings reveal several fruitful avenues for future work. We find that lottery wins do not impact higher-parity fertility but may improve child quality through increased overseas education. However, other quality measures like health status or child expenditures are unavailable in our administrative data. Understanding how cash windfalls affect the trade-off between child quantity and quality is an interesting research question for the future.

\nocite{*}

\newpage
\section*{Tables}
\pdfbookmark[-1]{Tables}{Tables}

\begin{center}
 \begin{threeparttable}
 \setlength{\tabcolsep}{5mm}{}
 \linespread{0.8}
 \fontsize{8.5}{8.5pt}\selectfont
 \centering\footnotesize
 \caption{Descriptive Statistics for Lottery Winners and the Population}
 \label{tab.descriptive} 
 \begin{tabular}{@{}lccc@{}}
 \toprule
 & \begin{tabular}[c]{@{}c@{}}Lottery Winners\end{tabular} 
 & \begin{tabular}[c]{@{}c@{}}Population\end{tabular} 
 & \begin{tabular}[c]{@{}c@{}}Difference\end{tabular} \\
 \midrule \midrule
 \textit{Individual characteristics}  &  &  &  \\ 
 ~~Age & 31.898 & 31.355 & 0.543*** \\
 & (6.736) & (7.896) & [0.011] \\
 ~~Living in urban area & 0.687 & 0.693 & -0.006*** \\
 & (0.464) & (0.461) & [0.001] \\
 ~~Female & 0.517 & 0.499 & 0.018*** \\
 & (0.500) & (0.500) & [0.001] \\
 ~~Married & 0.462 & 0.411 & 0.051*** \\
 & (0.499) & (0.492) & [0.001] \\
 ~~Winner's Employment & 0.748 & 0.694 & 0.054*** \\
 & (0.434) & (0.461) & [0.001] \\
 ~~Winner's Earnings  & 289.511 & 285.594 & 3.917*** \\
 & (394.448) & (546.489) & [0.640] \\
 ~~Winner's Income & 308.441 & 308.035 & 0.406 \\
 & (445.269) & (656.776) & [0.725] \\
 ~~Winner's Assets  & 2,041.302 & 2,320.071 & -278.769*** \\
 & (8,702.125) & (13,292.058) & [14.207] \\
 ~~Winner's Liquid Assets  & 612.436 & 709.105 & -96.669*** \\
 & (4,791.012) & (7,938.590) & [7.876] \\
~~Winner's Savings & 248.280 & 292.212 & -43.932*** \\
 & (1,155.056) & (1,390.573) & [1.858] \\
~~Household Earnings  & 490.417 & 458.116 & 32.301*** \\
 & (655.644) & (869.859) & [1.060] \\
 ~~Household Income  & 524.138 & 497.1978 & 26.940*** \\
 & (731.839) & (1,343.546) & [1.215] \\
 ~~Household Assets  & 3,846.935 & 4,165.627 & -318.692*** \\
 & (13,985.338) & (41,404.719) & [25.171] \\
 ~~Household Liquid Assets  & 1,065.459 & 1,209.224 & -143.765*** \\
 & (6,839.653) & (38,197.793) & [15.657] \\
 ~~Household Savings & 421.149 & 478.295 & -57.146*** \\
 & (1,688.944) & (2,440.002) & [2.746] \\
 \midrule
 \textit{Fertility variables} &  &  &  \\
 ~~Cumulative Number of Children & 0.881 & 0.817 & 0.064*** \\
 & (1.102) & (1.105) & [0.002] \\
 ~~Gave Birth in $s - 1$ & 0.044 & 0.034 & 0.010*** \\
 & (0.205) & (0.180) & [0.000] \\
 ~~Gave Birth in $s - 2$ & 0.045 & 0.034 & 0.011*** \\
 & (0.206) & (0.181) & [0.000] \\
 ~~Gave Birth in $s - 3$ & 0.046 & 0.035 & 0.011*** \\
 & (0.209) & (0.185) & [0.000] \\
 \midrule \midrule
 \# of Observations & 406,963 & 11,205,868 & \\
 \bottomrule
 \end{tabular}
 \begin{tablenotes}
 \fontsize{8}{8pt}\selectfont
Note: We utilize the all individuals aged 20-44 from 2007-2012 to construct population data. For each individual, we randomly assign one year between 2007-2012 as a placebo "winning year." We then use their individual characteristics from the year prior to this randomly assigned placebo winning year in our analysis. Urban areas refer to the 6 largest cities in Taiwan with special municipality status: Taipei City, New Taipei City, Taoyuan City, Taichung City, Tainan City, and Kaohsiung City. These cities have the largest populations in Taiwan. Employment is defined as having positive annual labor earnings. Annual earnings are defined as the sum of annual wage income, business income, and professional income. Annual income is defined as the sum of annual labor earnings plus other annual income sources like interest, rents, farming, pensions etc, excluding lottery winnings. Assets are defined as the sum of real estate value, financial assets, and stocks, minus mortgage debt. Liquid assets are defined as the sum of financial assets and stocks. All monetary values like earnings, income, assets and liquid assets are measured in thousand New Taiwan Dollars (NT\$)  and adjusted to 2016 NT\$ levels (1 NT\$ $\approx$ 0.033 US\$ in 2016). More details on the construction of asset data can be found in Appendix \ref{app: w_data}. Standard deviations are in parentheses, and standard errors are in brackets. *** significant at the 1 percent level, ** significant at the 5 percent level, and * significant at the 10 percent level.
 \end{tablenotes}
 \end{threeparttable}
\end{center}

\begin{center}
 \begin{threeparttable}
 \linespread{1.5}
 \fontsize{8.5}{8.5pt}\selectfont
 \centering\footnotesize
 \caption{Effect of a Ten Million NT\$ Lottery Prize on Fertility}\label{tab.main}
 \begin{tabular}{@{}lcccccc@{}}
 \toprule
 Dependent Variable: &\multicolumn{6}{c}{Number of Cumulative Children} \\
 \cmidrule(l){2-7}
 & (1) & (2) & (3) & (4) & (5) & (6) \\
 \midrule \midrule
 $Current_i \times Prize_i \times \mathbf{I}[t=L_{i}+6]$ & 0.051** & 0.045** & 0.046** & 0.047** & 0.048** & 0.048** \\
 & (0.020) & (0.019) & (0.019) & (0.019) & (0.019) & (0.019) \\ \\
 Baseline trend & \multicolumn{6}{c}{0.321} \\
 Observations & \multicolumn{6}{c}{4,069,630} \\
 \midrule \midrule
 Basic controls & $\surd$ & $\surd$ & $\surd$ & $\surd$ & $\surd$ & $\surd$ \\
 Age fixed effect & & $\surd$ & $\surd$ & $\surd$ & $\surd$ & $\surd$ \\
 Year fixed effect & & & $\surd$ & $\surd$ & $\surd$ & $\surd$ \\
 Individual characteristics & & & & $\surd$ & $\surd$ & $\surd$ \\
  Pre-treatment fertility & & & & & $\surd$ & $\surd$ \\
 Pre-treatment lottery winnings & & & & & & $\surd$ \\
 \bottomrule
 \end{tabular}
 \begin{tablenotes}
 \fontsize{8}{8pt}\selectfont
Note: This table reports estimated coefficients of $Current_i \times Prize_i \times \mathbf{I}[t=L_{i}+6]$ in Equation (\ref{child_event}), which stands for the effect of 10 Million NT\$ lottery wins on fertility at the sixth year following the receipt of cash windfalls. 
The outcome of interest is the cumulative number of children that winner $i$ has by the end of the sixth year after a lottery win. The baseline trend is the change in the cumulative number of children for the future winner between one year before and six years after the placebo lottery-winning year.
Column (1) includes the amount of lottery prize, a full set of event time dummies, the interaction terms between the lottery prize and even time dummies, and the full interactions between $current$ (a dummy indicating a current winner) and the above variables. Column (2) further includes the age fixed effect. Column (3) further includes the calendar year fixed effects. Column (4) includes pre-determined covariates: a set of dummies indicating cities/counties of residence, a dummy indicating the winner was married, a dummy indicating the winner or her spouse was employed, average household earnings per capita (evenly divided between spouses if married), average household income per capita (evenly divided between spouses if married), average household wealth per capita (evenly divided between spouses if married). Note that these covariates are measured in the year right before the lottery-winning year. Column (5) controls for the outcomes variable (the cumulative number of children) in the year right before the lottery-winning year. Column (6) includes the previous lottery winning record (the prize amount won in one, two, and three previous to the lottery winning year). Standard errors are clustered at the winner level and reported in parentheses.
 *** significant at the 1 percent level,
 ** significant at the 5 percent level, and
 * significant at the 10 percent level.
 \end{tablenotes}
 \end{threeparttable}
\end{center}

\begin{landscape}
	
\begin{center}
\linespread{1.2}
\begin{threeparttable}
\fontsize{8.5}{8.5pt}\selectfont
\centering
\caption{Robustness Checks}\label{tab.robust}
\begin{tabular}{@{}lccccccccc@{}}
\toprule
Dependent Variable: &\multicolumn{9}{c}{Number of Cumulative Children} \\
\cmidrule(){2-10}
& (1) & (2) & (3) & (4) & (5) & (6) & (7) & (8) & (9) \\
& \multicolumn{2}{c}{Minimum Win Amounts} & \multicolumn{2}{c}{Maximum Win Amounts} 
& \multirow{2}{*}{\begin{tabular}[c]{@{}c@{}} Population\\Weighted \end{tabular}}
& \multirow{2}{*}{\begin{tabular}[c]{@{}c@{}} Alternative\\Age FE \end{tabular}}
& \multirow{2}{*}{\begin{tabular}[c]{@{}c@{}} Individual\\FE \end{tabular}}
& \multirow{2}{*}{\begin{tabular}[c]{@{}c@{}} Single\\Event \end{tabular}}
& \multirow{2}{*}{\begin{tabular}[c]{@{}c@{}} Cohort-\\by-Cohort \end{tabular}}\\
\cmidrule(r){2-3} \cmidrule(l){4-5}
& 30K & 2K & 80M & 120M & & & & & \\
\midrule \midrule
$Current_i \times Prize_i \times \mathbf{I}[t=L_i+6]$ & 0.041** & 0.043** & 0.069*** & 0.041** & 0.042** & 0.050*** & 0.045** & 0.050** & 0.046** \\
& (0.017) & (0.020) & (0.026) & (0.017) & (0.018) & (0.019) & (0.018) & (0.023) & [0.022] \\
\\
\midrule
Baseline trend & 0.295 & 0.320 & 0.321 & 0.321 & 0.327 & 0.321 & 0.321 & 0.323 & 0.321 \\
Observations & 968,240 & 12,686,170 & 4,069,360  & 4,069,680  & 4,069,630 & 4,069,624 & 4,069,630 & 3,408,390 & 4,069,630 \\
\bottomrule
\end{tabular}
\begin{tablenotes}
\fontsize{8}{8pt}\selectfont
Note: This table reports estimated coefficients of $Current_i \times Prize_i \times \mathbf{I}[t=L_{i}+6]$ in Equation (\ref{child_event}), which stands for the effect of 10 Million NT\$ lottery wins on fertility at the sixth year following the receipt of cash windfalls. 
The outcome of interest is the cumulative number of children that winner $i$ has by the end of the sixth year after a lottery win. The baseline trend is the change in the cumulative number of children for the future winner between one year before and six years after the placebo lottery-winning year. 
Columns (1) and (2) report the estimate using different thresholds of ``minimum prizes,''-- the threshold we use to exclude observations---less than 30 thousand NT\$ (Column (1)) or less than 2 thousand NT\$ (Column (2)). 
Columns (3) and (4) report the estimate using different thresholds of ``maximum prizes,''---the threshold we use to exclude observations---above 80 Million NT\$ (Column (3)) or above 120 Million NT\$ (Column (4)). 
Column (5) reports the estimate based on re-weighting the lottery winner sample to make these characteristics similar to those of the general population in Taiwan. 
Column (6) includes both male and female age fixed effects. For winners with no spouse, we include a dummy indicating a missing spouse (either male or female). 
Column (7) includes individual (winner) fixed effects to account for any individual time-invariant unobserved factors.
Column (8) restricts samples to those who only win a lottery prize once in the sample period (2004 to 2018).
Column (9) estimates the lottery impact within each cohort ([placebo] lottery winning years) separately and the average effect weighted by the number of observations in each cohort. The standard error is obtained by 1,000 times bootstrapping (re-sampling within replacement within the cohorts [winning years] cluster).
Standard errors are clustered at the winner level and reported in parentheses. Bootstrapping standard errors are reported in squared brackets. \\
*** significant at the 1 percent level,
** significant at the 5 percent level, and
* significant at the 10 percent level.
\end{tablenotes}
\end{threeparttable}
\end{center}
\end{landscape}

\begin{center}
	\linespread{1.2}
	\begin{threeparttable}
		\fontsize{9.5}{9.5pt}\selectfont
		\centering
		\caption{Subgroup Analysis---By Amount of Lottery Prize}\label{tab.prizegrp}
		\begin{tabular}{@{}lccccc@{}}
			\toprule
			Dependent Variable: & \multicolumn{5}{c}{Number of Cumulative Children} \\ 
			\cmidrule(){2-6}
			& (1) & (2) & (3) & (4) & (5) \\
			& 10K--50K & 50K--500K & 500K--5M & 5M--100M & $\geq$100M \\
			\midrule \midrule
			$Current_i \times Large_i \times \mathbf{I}[t=L_i+6]$ & 0.008* & 0.020*** & 0.035* & 0.100*** & 0.348*** \\
 			& (0.004) & (0.007) & (0.019) & (0.030) & (0.128) \\
			\midrule
			Observations & 3,583,850 & 2,222,540 & 1,850,850 & 1,815,990 & 1,802,290 \\
			\bottomrule
		\end{tabular}
		\begin{tablenotes}
			\fontsize{8}{8pt}\selectfont
			Note: This table reports the estimated coefficients of $Current_{i} \times Large_{i} \times \mathbf{I}[t=L_{i}+6]$ in Equation (\ref{child_event_step}), which stands for the effect of large lottery prize wins on fertility in the sixth year following the receipt of a cash windfall. 
			The outcome of interest is the cumulative number of children that winner $i$ has by the end of the sixth year after the lottery win.
			All regressions include the same set of covariates shown in Column (6) of Table \ref{tab.main}. Columns (1) to (5) define the large prize group based on winning amounts. All columns compare the large prize to the small prize---defined as those who won between 5 to 10 thousand NT\$.
			Column (1) defines the large prize as 10 to 50 thousand NT\$.
			Column (2) defines the large prize as 50 to 500 thousand NT\$.
			Column (3) defines the large prize as 500 thousand to 5 million NT\$.
			Column (4) defines the large prize as 5 to 100 million NT\$.
			Column (5) defines the large prize as above 100 million NT\$.
			Standard errors are clustered at the winner level and reported in parentheses. \\
			*** significant at the 1 percent level,
			** significant at the 5 percent level, and
			* significant at the 10 percent level.
		\end{tablenotes}
	\end{threeparttable}
\end{center}

\newpage

\begin{center}
\linespread{1.2}
\begin{threeparttable}
\fontsize{8.5}{8.5pt}\selectfont
\centering
\caption{Subgroup Analysis---By Financial Resources}\label{tab.financial}
\begin{tabular}{@{}lcccccc@{}}
\toprule
Dependent Variable: & \multicolumn{6}{c}{Cumulative Number of Children} \\ 
\cmidrule(){2-7}
& (1) & (2) & (3) & (4) & (5) & (6) \\
& \multicolumn{2}{c}{Deposit} & \multicolumn{2}{c}{Liquid Asset} & \multicolumn{2}{c}{Asset} \\
\cmidrule(r){2-3} \cmidrule(rl){4-5} \cmidrule(l){6-7}
& = 0 & > 0 & = 0 & > 0 & $\leq$ 0 & > 0 \\
\midrule \midrule
$Current_i \times Prize_i \times \mathbf{I}[t=L_i+6]$ & 0.066*** & -0.013 & 0.100*** & -0.015 & 0.095*** & 0.005 \\
 & (0.024) & (0.026) & (0.026) & (0.024) & (0.030) & (0.022) \\ \\
\midrule
Baseline Trend & 0.313 & 0.345 & 0.313 & 0.332 & 0.320 & 0.322 \\
 Observations & 3,048,200 & 1,021,430 & 2,259,800 & 1,809,830 & 1,983,250 & 2,086,380 \\
\bottomrule
\end{tabular}
\begin{tablenotes}
\fontsize{8}{8pt}\selectfont
Note: This table reports estimated coefficients of $Current_i \times Prize_i \times \mathbf{I}[t=L_{i}+6]$ in Equation (\ref{child_event}), which stands for the effect of 10 Million NT\$ lottery wins on fertility in the sixth year following the receipt of a cash windfall. 
The outcome of interest is the cumulative number of children that winner $i$ has by the end of the sixth year after the lottery win. The baseline trend is the change in the cumulative number of children for the future winner between one year before and six years after the placebo lottery-winning year.
All regressions include the same set of covariates shown in Column (6) of Table \ref{tab.main}. 
Columns (1) and (2) divide the sample into two groups based on whether the winner had any deposits one year previous to the (placebo) winning year.
Column (1) reports the estimate based on winners with no deposits. Column (2) reports the estimate based on winners having a positive deposit. 
Columns (3) and (4) divide the sample into two groups based on whether the winner had liquid assets one year previous to the (placebo) winning year. Liquid assets is defined as the sum of market values of stock and capital savings. 
Column (3) reports the estimate for winners with no liquid assets. Column (4) reports the estimate for winners having liquid assets.
Columns (5) and (6) divide the sample into two groups based on the winner's total assets one year previous to the (placebo) winning year. Total assets is defined as the sum of liquid assets and real estate less mortgage debt.
Column (5) reports the estimate for winners whose total assets are negative or zero. Column (6) reports the estimate for winners whose total assets are greater than zero.
Standard errors are clustered at the winner level and reported in parentheses. \\
*** significant at the 1 percent level,
** significant at the 5 percent level, and
* significant at the 10 percent level.
\end{tablenotes}
\end{threeparttable}
\end{center}

\newpage
\begin{center}
\linespread{1.2}
\begin{threeparttable}
\fontsize{8.5}{8.5pt}\selectfont
\centering
\caption{Subgroup Analysis---By Parenthood Status and Household Status}\label{tab.household}
\begin{tabular}{@{}lcccccc@{}}
\toprule
Dependent Variable: & \multicolumn{6}{c}{Cumulative Number of Children} \\ 
\cmidrule(){2-7}
& (1) & (2) & (3) & (4) & (5) & (6) \\
\multicolumn{1}{c}{} & \multicolumn{2}{c}{Parenthood Status} 
& \multicolumn{4}{c}{Household Status} \\
\cmidrule(r){2-3} \cmidrule(l){4-7} 
& w/o Child & w/ Child & Single & Couple 
& \begin{tabular}[c]{@{}c@{}} Couple\\w/o Child \end{tabular}
& \begin{tabular}[c]{@{}c@{}} Couple\\w/ Child \end{tabular} \\
\midrule
\midrule
$Current_i \times Prize_i \times \mathbf{I}[t=L_i+6]$  & 0.093*** & 0.002 & 0.063** & 0.022 & 0.357** & -0.001 \\
 & (0.032) & (0.014) & (0.029) & (0.020) & (0.156) & (0.016) \\
\\
\midrule
Baseline Trend & 0.405 & 0.203 & 0.295 & 0.356 & 1.070 & 0.207 \\
Observations & 2,194,590 & 1,875,040 & 2,188,290 & 1,881,340 & 298,740 & 1,582,600 \\
\bottomrule
\end{tabular}
\begin{tablenotes}
\fontsize{8}{8pt}\selectfont
Note: This table reports estimated coefficients of $Current_i \times Prize_i \times \mathbf{I}[t=L_{i}+6]$ in Equation (\ref{child_event}), which stands for the effect of 10 Million NT\$ lottery wins on fertility in the sixth year following the receipt of a cash windfall. 
The outcome of interest is the cumulative number of children that winner $i$ has by the end of the sixth year after the lottery win. The baseline trend is the change in the cumulative number of children for the future winner between one year before and six years after the placebo lottery-winning year.
All regressions include the same set of covariates shown in Column (6) of Table \ref{tab.main}. 
Columns (1) and (2) separate the sample into two groups based on the cumulative number of children before the winning year. 
Column (1) includes winners with no child before winning the lottery. 
Column (2) includes winners with at least one child before winning the lottery. 
Columns (3) to (6) separate households into four groups based on family types.
Column (3) includes winners who were unmarried before winning the lottery. 
Column (4) includes winners who were married before winning the lottery
Column (5) includes married winners without children before winning the lottery.
Column (6) includes married winners with children before winning the lottery. 
Standard errors are clustered at the winner level and reported in parentheses. \\
*** significant at the 1 percent level,
** significant at the 5 percent level, and
* significant at the 10 percent level.
\end{tablenotes}
\end{threeparttable}
\end{center}

\newpage

\begin{center}
\linespread{1.2}
\begin{threeparttable}
\fontsize{8.5}{8.5pt}\selectfont
\centering
\caption{Effect of a Ten Million NT\$ Lottery Prize on Children's College Attendance}\label{tab.education}
\begin{tabular}{@{}lccccccccc@{}}
\toprule
& (1) & (2) & (3) & (4) & (5) & (6) & (7) & (8) & (9) \\
\midrule \midrule
Dependent Variable: & \multicolumn{3}{c}{Ever Attend Any College} 
& \multicolumn{3}{c}{Ever Attend Domestic College} & \multicolumn{3}{c}{Ever Attend Overseas College} \\
\cmidrule(r){2-4} \cmidrule(rl){5-7} \cmidrule(l){8-10}
$Current_j \times Prize_j$ & -0.026 & -0.025 & -0.028 & -0.027 & -0.026 & -0.029 & 0.016* & 0.017* & 0.017* \\
 & (0.037) & (0.036) & (0.035) & (0.037) & (0.036) & (0.035) & (0.008) & (0.008) & (0.008) \\
\\
\midrule
Baseline mean & \multicolumn{3}{c}{0.734} & \multicolumn{3}{c}{0.727} & \multicolumn{3}{c}{0.014} \\
Observations & \multicolumn{3}{c}{80,661} & \multicolumn{3}{c}{80,661} & \multicolumn{3}{c}{80,661} \\
\midrule \midrule
Year fixed effect & $\surd$ & $\surd$ & $\surd$ & $\surd$ & $\surd$ & $\surd$ & $\surd$ & $\surd$ & $\surd$ \\
Cohort fixed effect & $\surd$ & $\surd$ & $\surd$ & $\surd$ & $\surd$ & $\surd$ & $\surd$ & $\surd$ & $\surd$ \\
Parental Control & & $\surd$ & $\surd$ & & $\surd$ & $\surd$ & & $\surd$ & $\surd$ \\
Child Control & & & $\surd$ & & & $\surd$ & & & $\surd$ \\
\bottomrule
\end{tabular}
\begin{tablenotes}
\fontsize{8}{8pt}\selectfont
Note: This table reports estimated coefficients of $Current_j \times Prize_j$ in Equation (\ref{child_q}), which stands for the effect of 10 Million NT\$ lottery wins on winners' children's college attendance. 
The outcomes of interest are dummies indicating the child ever attended any college (Columns (1) to (3)), domestic college (Columns (4) to (6)), or overseas college (Columns (7) to (9)) as of age 19. The baseline mean is the mean of the outcome variables for the future winners (those who won a lottery prize at a later period when their children were already greater than age 19). 
Columns (1), (4), and (7) include only basic DID variables---the amount of winnings, a dummy indicating the parent is a current winner, and the interaction term of the two ---, the child's cohort fixed effect, and the calendar year fixed effect. 
Columns (2), (5), and (8) further include winners' (parental) covariates---a set of dummies indicating cities/counties of residence, a dummy indicating the winner was married, a dummy indicating the winner or her spouse was employed, average household earnings per capita (evenly divided between spouses if married), average household income per capita (evenly divided between spouses if married), average household wealth per capita (evenly divided between spouses if married), the cumulative number of children in the year right before the lottery-winning year, and the prize amount won in one, two, and three years prior to the lottery-winning year.
Columns (3), (6), and (9) further include children's covariates---gender, birthplace, birth order, and birth month.
Standard errors are clustered at the winner level and reported in parentheses. \\
*** significant at the 1 percent level,
** significant at the 5 percent level, and
* significant at the 10 percent level.
\end{tablenotes}
\end{threeparttable}
\end{center}

\begin{center}
 \begin{threeparttable}
 \linespread{1.5}
 \fontsize{8.5}{8.5pt}\selectfont
 \centering\footnotesize
 \caption{Effect of a Ten Million NT\$ Lottery Prize on Marriage}\label{tab.married}
 \begin{tabular}{@{}lcccccc@{}}
 \toprule
 Dependent Variable: &\multicolumn{6}{c}{Ever Getting Married} \\
 \cmidrule(l){2-7}
 & (1) & (2) & (3) & (4) & (5) & (6) \\
 \midrule \midrule
 $Current_i \times Prize_i \times \mathbf{I}[t=L_{i}+6]$ &0.022** & 0.023** & 0.023** & 0.023** & 0.023** & 0.023** \\
    & (0.011) & (0.011) & (0.011) & (0.011) & (0.011) & (0.011)  \\ \\
 Baseline trend & \multicolumn{6}{c}{0.163} \\
 Observations & \multicolumn{6}{c}{4,069,630} \\
 \midrule \midrule
 Basic controls & $\surd$ & $\surd$ & $\surd$ & $\surd$ & $\surd$ & $\surd$ \\
 Age fixed effect & & $\surd$ & $\surd$ & $\surd$ & $\surd$ & $\surd$ \\
 Year fixed effect & & & $\surd$ & $\surd$ & $\surd$ & $\surd$ \\
 Individual characteristics & & & & $\surd$ & $\surd$ & $\surd$ \\
 Pre-treatment fertility & & & & & $\surd$ & $\surd$ \\
 Pre-treatment lottery winnings & & & & & & $\surd$ \\
 \bottomrule
 \end{tabular}
 \begin{tablenotes}
 \fontsize{8}{8pt}\selectfont
Note: This table reports estimated coefficients of $Current_i \times Prize_i \times \mathbf{I}[t=L_{i}+6]$ in Equation (\ref{child_event}), which stands for the effect of 10 Million NT\$ lottery wins on fertility in the sixth year following the receipt of a cash windfall. 
The outcome of interest is ever getting married by the end of the sixth year after the lottery win. The baseline trend is the change in the proportion of individuals ever married for the future winner between one year before and six years after the placebo lottery-winning year.
Column (1) includes the amount of winnings, a full set of event time dummies, the interaction terms between lottery prize and even time dummies, and the full interactions between $current$ (a dummy indicates current winner) and the above variables. Column (2) further includes the age fixed effect. Column (3) further includes the calendar year fixed effects. Column (4) includes pre-determined covariates, i.e., a set of dummies indicating cities/counties of residence, a dummy indicating the winner was married, a dummy indicating the winner or her spouse was employed, average household earnings per capita (evenly divided between spouses if married), average household income per capita (evenly divided between spouses if married), average household wealth per capita (evenly divided between spouses if married). Note that these covariates are measured in the year right before the lottery-winning year. Column (5) controls for the pre-treatment fertility (the cumulative number of children) in the year right before the lottery-winning year. Column (6) includes the previous lottery-winning record (the prize amount won in one, two, and three years previous to the lottery winning year). Standard errors are clustered at the winner level and reported in parentheses.
 *** significant at the 1 percent level,
 ** significant at the 5 percent level, and
 * significant at the 10 percent level.
 \end{tablenotes}
 \end{threeparttable}
\end{center}

\newpage
\section*{Figures}
\pdfbookmark[-1]{Figures}{Figures}

\newpage


\newpage

\newpage

\newpage

\newpage
\section*{Online Appendix: For Online Publication}
\pdfbookmark[-1]{Online Appendix}{Online Appendix}

\appendix

\begin{center}
	\setlength{\tabcolsep}{7mm}{
		\Large   
		\begin{tabularx}
			{\linewidth}{l >{\raggedright\arraybackslash}X >{\raggedright\arraybackslash}X}
			Section A & \nameref{app: add_lottery}  \\	
			Section B & \nameref{app: w_data}  \\	
			
			Section C & \nameref{app: add_t_f} \\
		\end{tabularx}
	}
\end{center}

\newpage
\setcounter{table}{0}
\renewcommand{\thetable}{A\arabic{table}}
\setcounter{figure}{0}
\renewcommand{\thefigure}{A\arabic{figure}}

\section{Lottery Games in Taiwan}\label{app: add_lottery}

\subsection{Public Welfare Lottery}
The Taiwanese government initiated the Public Welfare Lottery in 1999. The purpose of the lottery was to improve social welfare by creating job opportunities for the disabled, native aborigines, and single-parent families to sell tickets. The government uses revenue from selling lottery tickets to support its social welfare program. There were three main types of Public Welfare Lottery in our sample period: (1) Computer-drawn games, (2) scratchcard games, and (3) Keno games.

\begin{figure}[H]
	\centering
	\caption{Computer-Drawn Game - Lott 6/49}\label{loto1}
	\begin{subfigure}[t]{0.475\textwidth}
		\caption{Purchase Sheet} \label{loto1a}
		\includegraphics[width=1\linewidth]{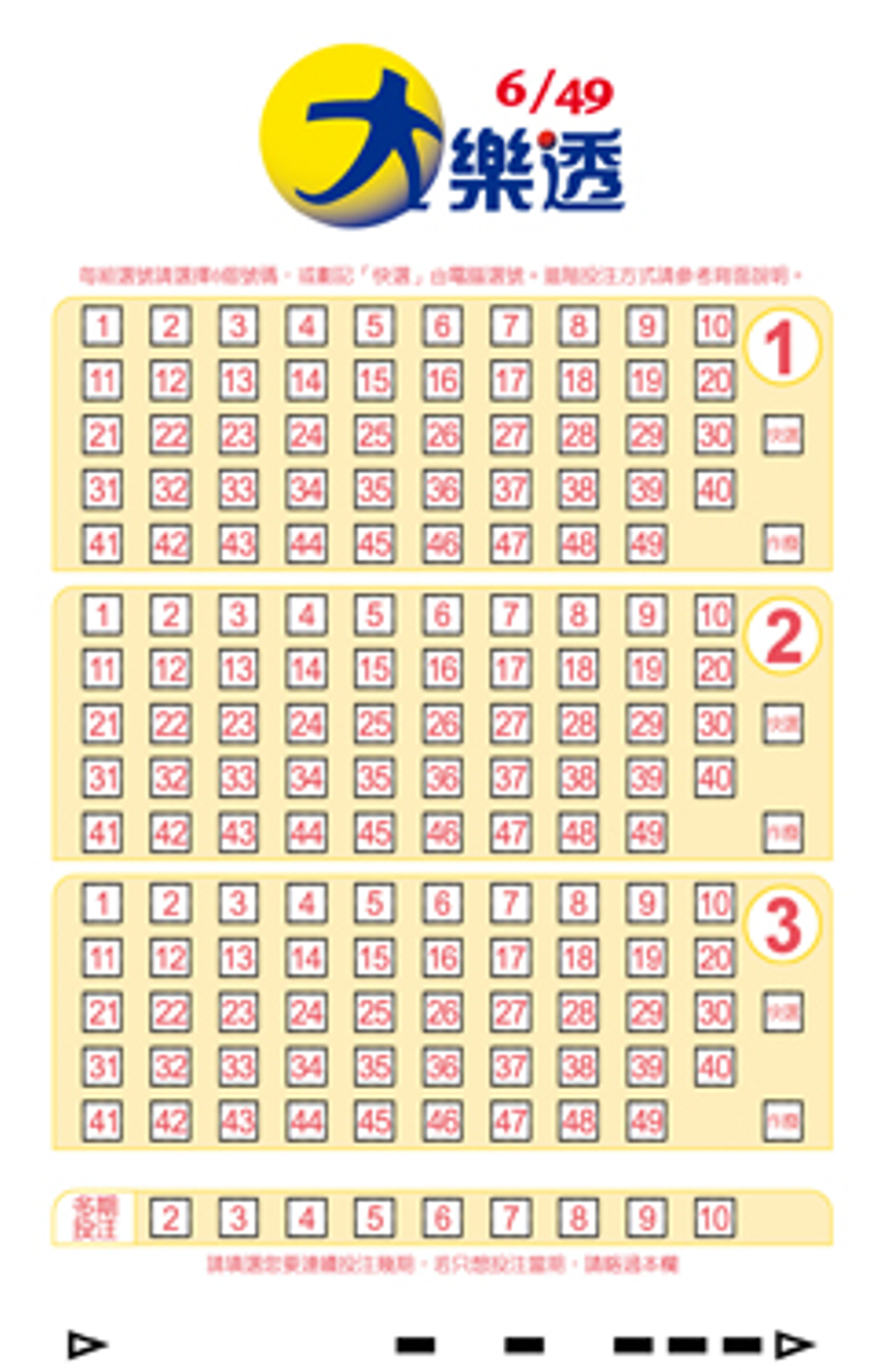}
	\end{subfigure}
	\begin{subfigure}[t]{0.475\textwidth}
		\caption{Purchase Receipt} \label{loto1b}
		\includegraphics[width=1\linewidth]{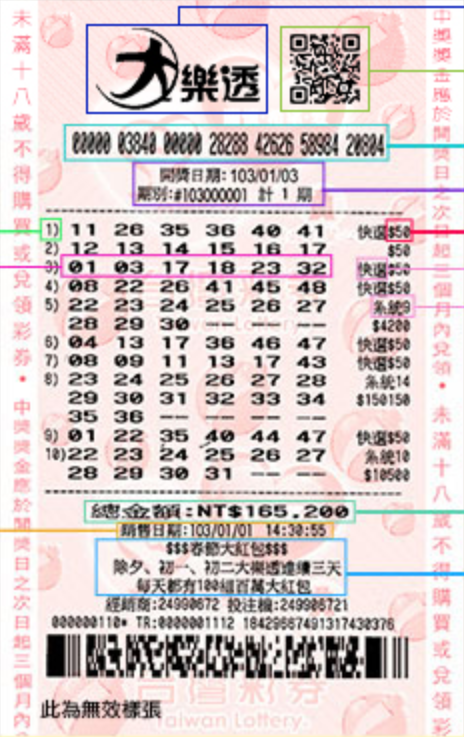}
	\end{subfigure}

	\fontsize{10}{10pt}\selectfont
	\flushleft
	\emph{Notes:} Figure \ref{loto1a} displays the  Lott 6/49 purchase sheet. Each sheet has multiple sections. Players choose six numbers from section one for one bet. If players want to have more than one bet, they can repeat the same process in other sections. Players also can choose the right column below the section number to let the betting machine choose six numbers randomly. After submitting the purchase sheet to the betting shop, the player receives a receipt, as displayed in Figure \ref{loto1b}, which can be used to redeem the prize. \\
	Figure source: Taiwan Lottery Website. \url{https://www.taiwanlottery.com.tw/Lotto649/index.asp}.
\end{figure}

In this section, we present demos of these lottery tickets or cards. For a computer-drawn game, players need to select a set of numbers. The lottery agency regularly announces the prize numbers drawn by the computer (e.g., twice a week). For example, Lotto 6/49 is one of the most popular computer-drawn games in Taiwan, a ticket for which is presented in Figure \ref{loto1}. Players choose six numbers (1-49) at a cost of 50 NT\$ per bet. The prize amount depends on how many numbers match, and the jackpot is hit if all six numbers are matched. The jackpot keeps growing until someone wins. 

\begin{figure}[H]
	\centering
	\caption{Scratched Game}\label{scratched}
	\begin{subfigure}[t]{0.475\textwidth}
		\caption{Unscratched} \label{scratcheda}
		\includegraphics[width=1\linewidth]{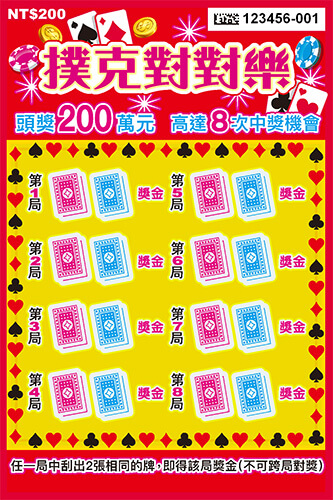}
	\end{subfigure}
	\begin{subfigure}[t]{0.475\textwidth}
		\caption{Scratched} \label{scratchedb}
		\includegraphics[width=1\linewidth]{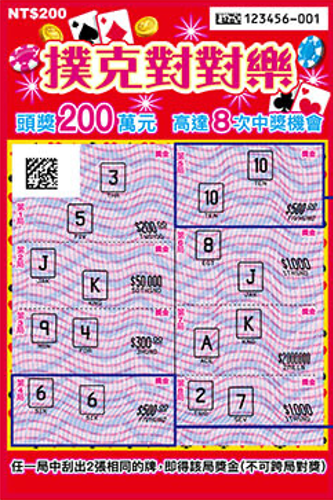}
	\end{subfigure}	
	
	\fontsize{10}{10pt}\selectfont
	\flushleft
	\emph{Notes:} This figure displays one of the famous scratchcard games. Figure \ref{scratcheda} displays the unscratched card, which has eight sets of games. Players need to scratch the card and match the numbers in each set to win specific prizes. As shown in Figure \ref{scratchedb}, the numbers inside the blue shape matched each other. Hence, the player won the prize as shown in the shape. \\
	Figure source: Taiwan Lottery Website. \url{https://www.taiwanlottery.com.tw/instant/instant_games_details_4573.asp}.
\end{figure}

\begin{figure}[H]
	\centering
	\caption{Keno Game}\label{bingobingo}
	\begin{subfigure}[t]{0.475\textwidth}
		\caption{Purchase Sheet} \label{bingobingoa}
		\includegraphics[width=1\linewidth]{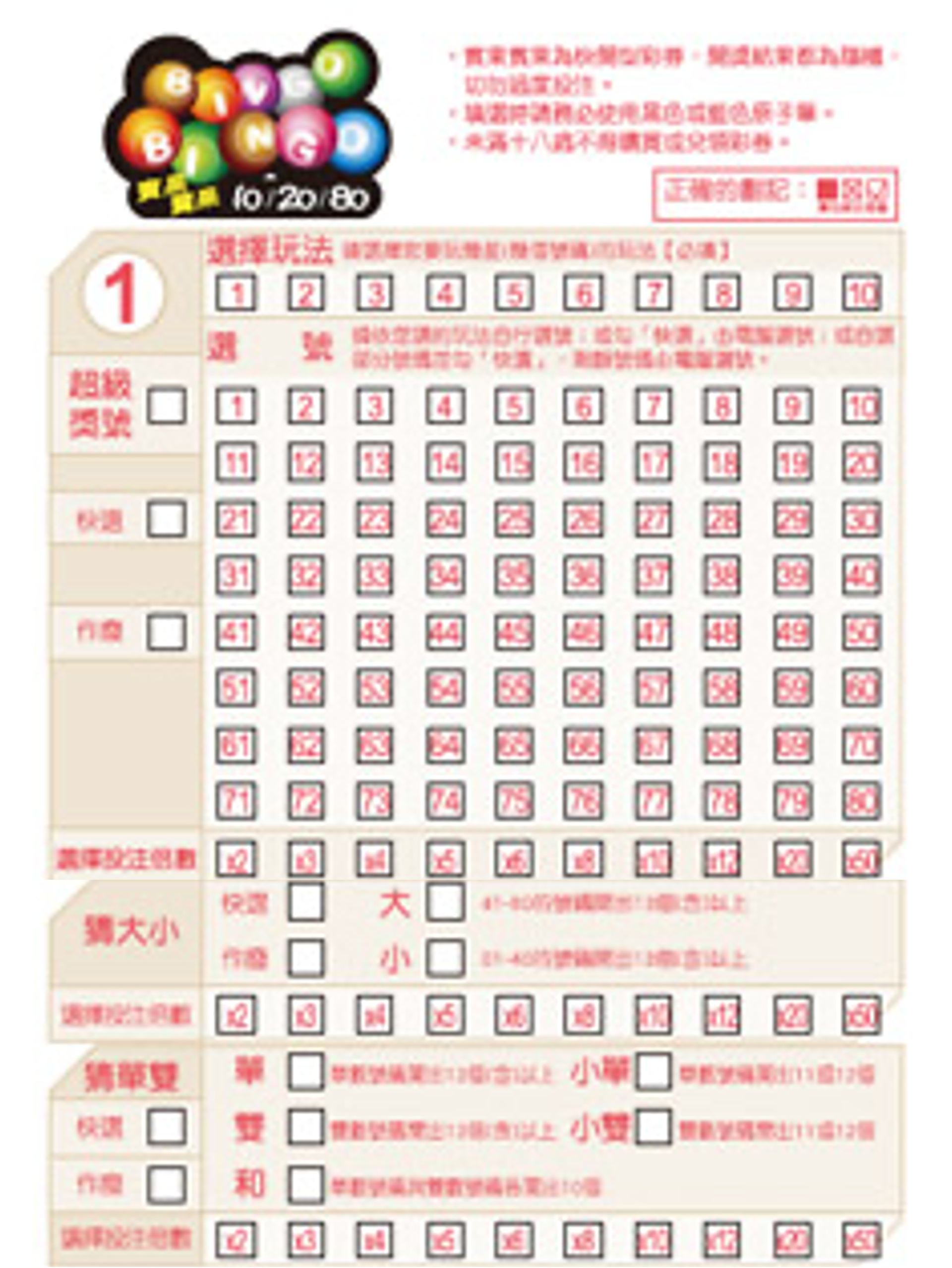}
	\end{subfigure}
	\begin{subfigure}[t]{0.475\textwidth}
		\caption{Purchase Receipt} \label{bingobingob}
		\includegraphics[width=1\linewidth]{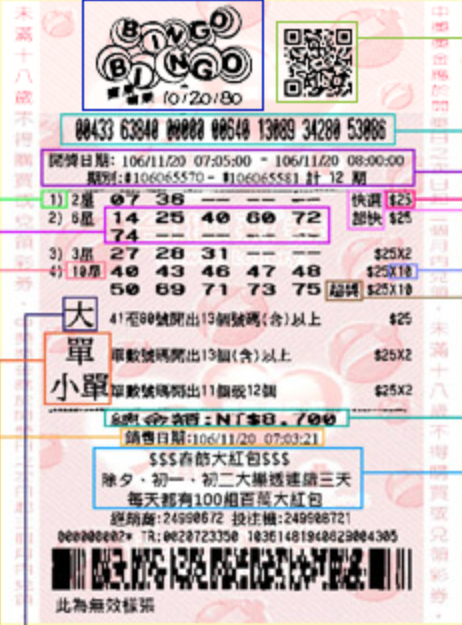}
	\end{subfigure}	
	
	\fontsize{10}{10pt}\selectfont
	\flushleft
	\emph{Notes:} Figure \ref{bingobingoa} displays a Keno game purchase sheet. Players first choose one of ten gameplays in the first row, then choose 20 numbers from 1 to 80. They can also bet whether the numbers will be high or low, even or odds in the bottom panel. After submitting the purchase sheet to the betting shop, the player receives a receipt as shown in Figure \ref{bingobingob}, which can be used to redeem a prize. The prizes are different according to the gameplay and how many numbers a player matches. \\
	Figure source: Taiwan Lottery Website. \url{https://www.taiwanlottery.com.tw/BINGOBINGO/index.asp}.
\end{figure}

Scratchcard games usually require a player to scratch away numbers or symbols to reveal specific prizes. Figure \ref{scratched} shows a typical type of scratchcard game. In Keno games, players need to select a set of numbers and game types. Figure \ref{bingobingo} shows a ticket for a Keno game. The common rule of Keno games is that a player chooses one of ten gameplays and then selects 20 numbers, ranging from 1 to 80. Payouts are different depending on the gameplay and the numbers a player chooses. 

\subsection{Taiwan Receipt Lottery}
\begin{figure}[H]
	\centering
	\caption{Taiwan Receipt Lottery}\label{receipt}
	\includegraphics[width=0.5\linewidth]{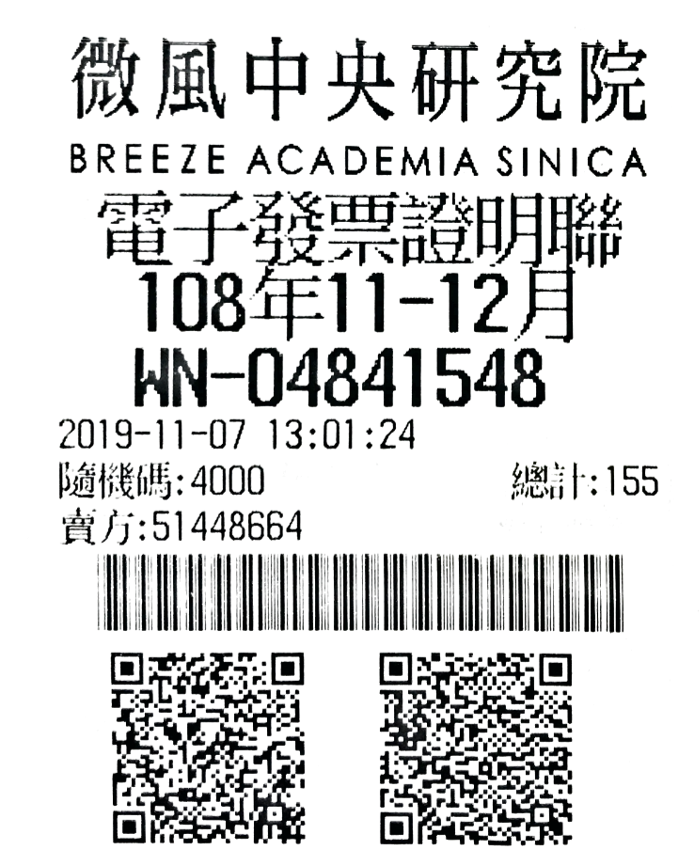} \\
	\fontsize{10}{10pt}\selectfont
	\flushleft
	\emph{Notes:} This figure displays an example from the Receipt Lottery, which contains 8 numbers (04841548).
\end{figure}

The Taiwan Receipt Lottery started in 1950. Its purpose is to ensure that consumers ask for receipts from sellers and therefore prevent tax evasion. Having purchased any goods or services, including paying electricity or telephone bills, the consumer receives an invoice with an eight-digit number printed along the top. Figure \ref{receipt} shows a sample of an invoice. The government then draws and announces the winning numbers bi-monthly. No matter the amount paid for an item, each receipt invoice has an even chance of winning the lottery by matching the drawn number. Table \ref{wn} presents the game rules and prize amounts, ranging from 200 NT\$ (about 6.7 US\$) to 2 million NT\$ (about 67 thousand US\$) before 2011. The largest prize rose to 10 million NT\$ (about 333 thousand US\$) in 2011.

\newpage
\begin{table}[H]
	\linespread{1}
	\fontsize{8.5}{8.5pt}\selectfont
	\centering\footnotesize
	\caption{Rules for the Taiwan Receipt Lottery}\label{wn}
	\begin{tabular}{lcl}
		\toprule
		\multicolumn{2}{c}{Prizes (in TWD)}& Matching Winning Numbers \\
		\midrule
		Special Prize & 10 million & all 8 digits from the special prize number \\
		Grand Prize & 2 million &  all 8 digits from the grand prize number\\
		First Prize  & 200,000  &  all 8 digits from any of the First Prize numbers\\
		Second Prize & 40,000  &  the last 7 digits from any of the First Prize numbers\\
		Third Prize & 10,000  &  the last 6 digits from any of the First Prize numbers\\
		Fourth Prize & 4,000   &  the last 5 digits from any of the First Prize numbers\\
		Fifth Prize & 1,000 &  the last 4 digits from any of the First Prize numbers\\
		Sixth Prize  & 200  &  the last 3 digits from any of the First Prize numbers\\
		Additional Sixth Prize  & 200  &  the last 3 digits from the Additional Sixth Prize number(s)\\
		\bottomrule
	\end{tabular}\\ [0.2cm]
	\begin{minipage}{0.92\textwidth}
		\fontsize{9}{9pt}\selectfont
		Note: This table displays the rules for the Taiwan Receipt Lottery. People receive an entry, which contains 8 numbers (see Figure \ref{receipt}), when they purchase goods. They match these numbers on the receipt to the numbers randomly drawn by the Ministry of Finance every two months. 
	\end{minipage}
\end{table}%

\newpage

\newpage

\newpage

\newpage
\section{Construction of Individual Wealth Data}\label{app: w_data}
We construct individual wealth data using the following administrative records: (1) wealth registry; (2) income statement file; and (3) records on mortgage interest costs. The wealth register contains the third-party reported variables of financial and non-financial assets for all individuals in Taiwan. Financial assets include detailed information on end-of-year listed and unlisted stocks. The price of stocks is measured by the trading price at the ex-dividend date on the Taiwan Stock Exchange (TWSE) and the Taipei Exchange.\footnote{The Taipei Exchange is the stock exchange for listed companies in the Over-the-Counter (OTC) market and the emerging stock market. For those stocks with no information on ex-dividend date trading price, we use the closing price at the end of July instead; for those stocks that do not have a closing price at the end of July, we use the net asset value share instead.}
The stocks of unlisted companies are also included and priced by the net asset value share.\footnote{The net asset value of a company is defined as the total assets (including cash, saving, merchandise inventory, equipment, investments, etc.) and liabilities (including loan, accounts payable, pension reserves, etc.) as listed on the income return file of the company.}

Non-financial assets include real estate (lands and houses). The information includes areas, locations, and unique identification numbers.
The value of land and houses in wealth registers is measured by their assessed values, which are announced by the local government in Taiwan for tax purposes once per year, and it are considered much lower than trading prices in the market.\footnote{For example, the assessed value of a house is based on construction costs, depreciation, and location ranking adjustment.}
To bring the value of land and houses closer to the trading price in the market,
following the procedures in \citet{LethPetersen2010} and \citet{Boserup2016a}, we
multiply the assessed values of houses or land by the ratio of average trading prices to average assessed values at the township level.


However, the wealth register does have two limitations. First, bank deposits, bonds, and other assets in the money markets, such as short-term bills, are not included in the data.\footnote{Individual interest income of bonds and short-term bills is taxed separately with uniformly 10\%. They are common tools used to save individual income tax in Taiwan.}
Therefore, we estimate the value of these assets by using information on interest income from the income statement and a simple capitalization method \citep{Saez2016}. To start, we separately aggregate the interest income of deposits, bonds, and short-term bills.
Then, in order to construct the capitalization rate, we divide each aggregate interest income by the aggregate amount of assets (i.e., bank deposits, bonds, and short-term bills) reported in the Financial Statistics published by the Central Bank. Thus, $r_{jt}$ is the capitalization rate for asset $j$ in year $t$, defined as follows:

\renewcommand{\theequation}{B.\arabic{equation}}
    \setcounter{equation}{0}
    
\begin{align}
r_{jt}=\frac{\sum_i d_{ijt}}{W_{jt}},
\end{align}
where $d_{ijt}$ is the interest income for asset $j$ held by individual $i$ in year $t$, and $W_{jt}$ is the corresponding aggregate amount of the asset $j$ in year $t$. Finally, we can calculate each individual $i$'s capitalized assets $w_{ijt}$ by dividing interest income $d_{ijt}$ of asset $j$ by the corresponding capitalization rate $r_{jt}$.
\begin{align}
w_{ijt}=\frac{d_{ijt}}{r_{jt}},
\end{align}

Second, the wealth registry data lack information on debt. We use records on mortgage interest costs reported by third-party (i.e., banks) and the same capitalization method to impute the value of debt for each individual. According to \citet{lien2021wealth}, mortgages reported to the tax agency account for around 53\% of total debt in Taiwan.
Therefore, we think our wealth data should cover most debts held by the Taiwanese. Finally, one important reminder is that pensions and insurance are not included in our wealth data, which is a common drawback of administrative wealth data in Nordic countries. According to the National Wealth Report, pensions and insurances account for 17\% of individuals' total net wealth in 2014.\footnote{Retrieved from: \url{https://www.stat.gov.tw/public/Data/861393520GEYI9Z14.pdf}. Date of access: July 31, 2022.}

\newpage

\setcounter{table}{0}
\setcounter{figure}{0}
\renewcommand{\thetable}{C\arabic{table}}
\renewcommand{\thefigure}{C\arabic{figure}}
\section{Additional Tables and Figures}\label{app: add_t_f}

\begin{figure}[H]
	\centering
	\caption{The Relationship Between GDP Per Capita and the Total Fertility Rate}\label{fig.gdp_fertility}
	\includegraphics[width=0.8\linewidth]{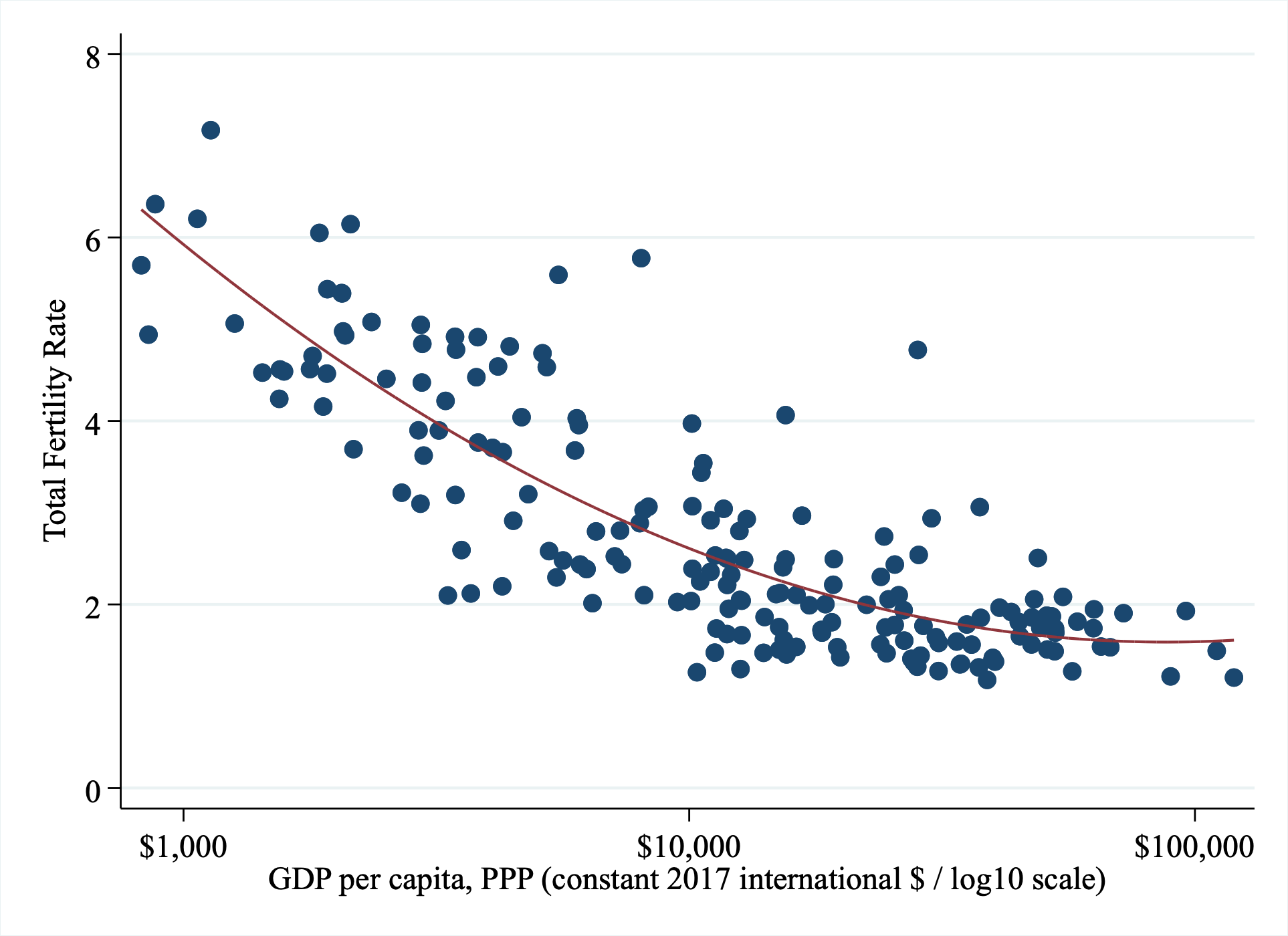}\\
	\fontsize{10}{10pt}\selectfont
	\flushleft
	\emph{Notes:} Each symbol stands for one country. The total fertility rate is defined as the number of children per 1,000 women. The data year is 2020. Data source: Our World in Data \citep{owidfertilityrate,owidgdp}.
\end{figure}

\begin{figure}[H]
	\centering
	\caption{Effect of a Ten Million NT\$ Lottery Prize on Fertility (Tracking 8 Years)} \label{fig.b3a8}	 
		\includegraphics[width=0.8\textwidth]{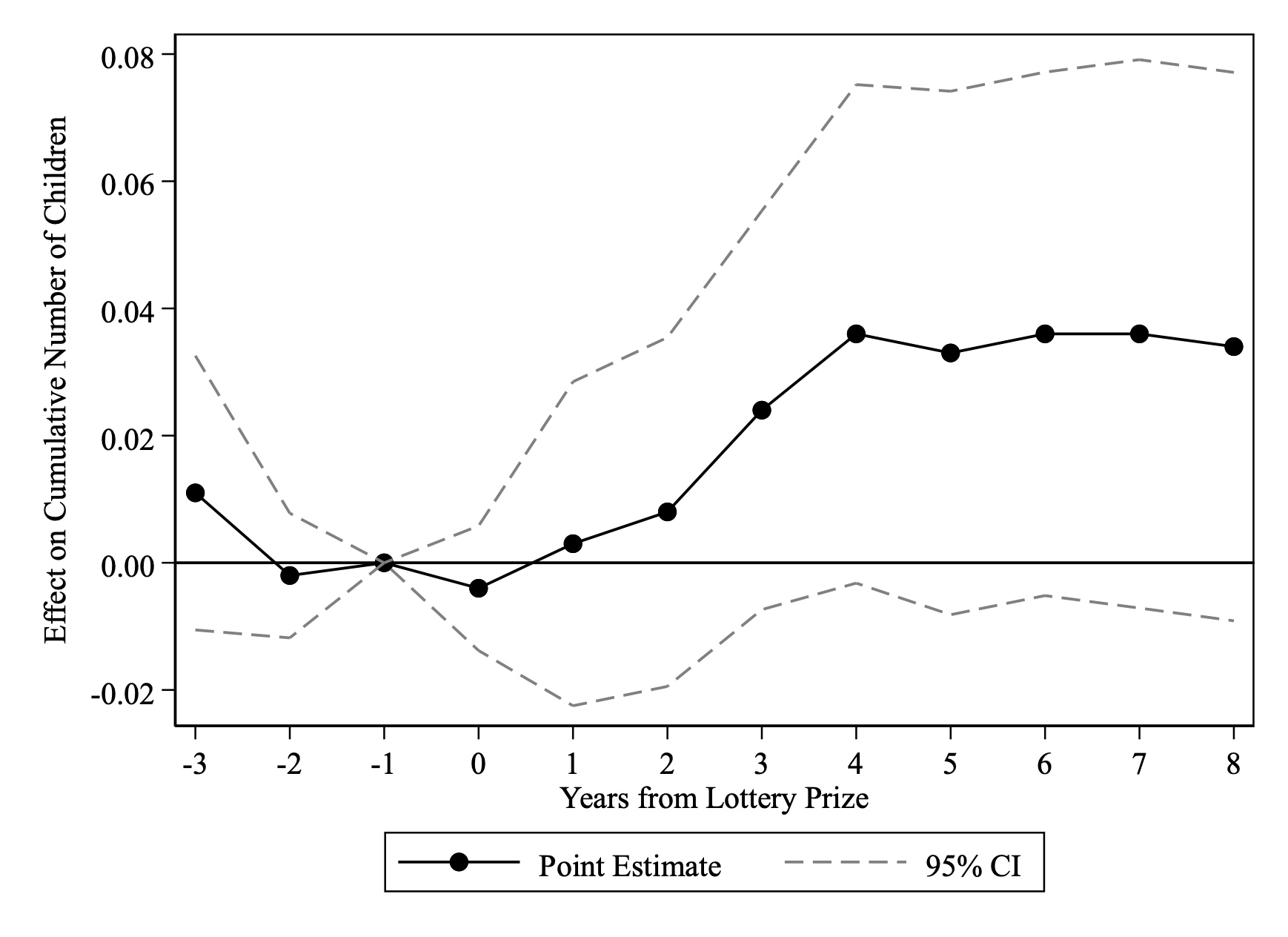}
	\fontsize{10}{10pt}\selectfont
	\flushleft
	\emph{Notes:} This figure displays the estimated coefficients of $Current_{i} \times Prize_{i} \times \mathbf{I}[t=L_{i}+s]$ from Equation (\ref{child_event}), but set $s$ = -3, -2,..., 8. The sample only consists of winners whose (placebo) winning years are from 2007 to 2010 (four cohorts). The outcome of interest is the cumulative number of children. The solid line denotes the point estimates. The dashed line denotes the 95\% confidence interval. The horizontal axis refers to the number of years from the (placebo) lottery-winning year.
\end{figure}

\newpage
\begin{center}
\linespread{1.5}
\begin{threeparttable}
\fontsize{10}{10pt}\selectfont
\centering
\caption{Distribution of a Lottery Prize}\label{tab.distribution}
\begin{tabular}{@{}llll@{}}
\toprule
 Prize Amount & Number of Winners 
 & \begin{tabular}[c]{@{}c@{}}Mean Win \\(Thousand NT\$)\end{tabular}  
 & \begin{tabular}[c]{@{}c@{}}Median Win \\(Thousand NT\$)\end{tabular}  \\
 \midrule \midrule
 \textbf{All Prizes} & & & \\
 5K--10K & 180,120 & 8 & 8 \\
 10K--50K & 178,265 & 20 & 17 \\
 50K--500K & 42,134 & 118 & 92 \\
 500K--5M & 4,965 & 1,334 & 812 \\
 5M--100M & 1,479 & 10,059 & 6,827 \\
 >100M$^\dagger$ & 109 & 377,099 & 281,663 \\
 \midrule
  \textbf{Public Welfare Lottery} & & & \\
 5K--10K & 96,275 & 7 & 7 \\
 10K--50K & 166,113 & 19 & 17 \\
 50K--500K & 41,173 & 117 & 90 \\
 500K--5M & 4,490 & 1,322 & 811 \\
 5M--100M & 1,275 & 10,379 & 6,726 \\
 >100M$^\dagger$ & 109 & 377,099 & 281,663 \\
 \midrule
  \textbf{Taiwan Receipt Lottery} & & & \\
 5K--10K & 95,862 & 8 & 8 \\
 10K--50K & 10,771 & 28 & 32 \\
 50K--500K & 918 & 164 & 162 \\
 500K--5M & 475 & 1,452 & 1,622 \\
 5M--100M & 204 & 8,061 & 8,000 \\
\bottomrule
\end{tabular}
\begin{tablenotes}
\fontsize{8}{8pt}\selectfont
Note: All prizes are after-tax amounts and adjusted with CPI, displayed in 2016 NT\$ (1 NT\$ $\approx$ 0.033 US\$). An individual can win both the Public Welfare Lottery and the Taiwan Receipt Lottery in a given year. Therefore, the sum of the head counts of two subcategories might exceed the total head counts. \\
$^\dagger$ Not included in the main sample.
\end{tablenotes}
\end{threeparttable}
\end{center}
	
\newpage
\begin{center}
\linespread{1.2}
\tabcolsep=1.5pt
\begin{threeparttable}
\fontsize{8.5}{8.5pt}\selectfont
\centering
\caption{Balance Test}\label{tab.balance}
\begin{tabular}{@{}lccccccccccc@{}}
\toprule
& (1) & (2) & (3) & (4) & (5) & (6) & (7) & (8) & (9) & (10) & (11) \\
\cmidrule(){2-12}
& \multicolumn{4}{c}{} & \multicolumn{5}{c}{Winner's Income \& Asset (10M)} & & \\
\cmidrule(){6-10}
\begin{tabular}[l]{@{}l@{}}Dependant \\Variable:\end{tabular} & Urban & Female & Married & Employed 
& \begin{tabular}[c]{@{}c@{}}Earnings\end{tabular} 
& \begin{tabular}[c]{@{}c@{}}Income\end{tabular} 
& \begin{tabular}[c]{@{}c@{}}Assets\end{tabular} 
& \begin{tabular}[c]{@{}c@{}}Liquid\\Assets\end{tabular} 
& \begin{tabular}[c]{@{}c@{}}Savings\end{tabular} 
& \begin{tabular}[c]{@{}c@{}}Number\\of\\Children\end{tabular} 
& \begin{tabular}[c]{@{}c@{}}Total\\Prize\\(1K)\end{tabular} \\
\midrule \midrule
 \multicolumn{12}{l}{\textbf{Panel A: Current Winners}} \\
 $Prize_i$ & 0.027*** & -0.052*** & -0.021** & -0.020* & 0.000 & 0.001 & 0.005 & -0.004 & -0.002* & -0.035 & -0.089*** \\
 & (0.008) & (0.012) & (0.010) & (0.010) & (0.001) & (0.001) & (0.010) & (0.003) & (0.001) & (0.022) & (0.027) \\
 Observations & \multicolumn{11}{c}{222,955} \\
\midrule
 \multicolumn{12}{l}{\textbf{Panel B: Future Winners}} \\
 $Prize_i$ & 0.016 & -0.082*** & -0.014 & -0.003 & 0.001 & 0.001 & 0.015 & -0.002 & -0.002 & -0.024 & -0.058*** \\
 & (0.010) & (0.012) & (0.009) & (0.008) & (0.001) & (0.001) & (0.016) & (0.004) & (0.002) & (0.022) & (0.017) \\
 Observations & \multicolumn{11}{c}{184,008} \\
\midrule
 \multicolumn{12}{l}{\textbf{Panel C: Difference-in-differences}} \\
$Current_i \times Prize_i$ & 0.011 & 0.029* & -0.008 & -0.016 & -0.000 & -0.000 & -0.011 & -0.002 & -0.000 & -0.012 & -0.023 \\
 & (0.013) & (0.017) & (0.014) & (0.013) & (0.001) & (0.001) & (0.019) & (0.005) & (0.003) & (0.031) & (0.032) \\
Observations & \multicolumn{11}{c}{406,963} \\
\midrule \midrule
Baseline mean & 0.682 & 0.545 & 0.424 & 0.827 & 0.027 & 0.029 & 0.190 & 0.060 & 0.025 & 0.791 & 0.270 \\
\bottomrule
\end{tabular}
\begin{tablenotes}
\fontsize{8}{8pt}\selectfont
Note: Panels A and B report estimated coefficients of $\beta$ from the equation $X_i=\beta \cdot Prize_i + a_{i} + \varepsilon_{i}$. Panel A only includes current winners, and Panel B only includes future winners. Panel C reports estimated coefficients of $\beta_{3}$ from Equation $X_i=\beta_{1} Current_i +\beta_{2} Prize_i + \beta_{3} Current_i \times Prize_i + a_{i} + \varepsilon_{i}$. The outcomes of interest are winners' characteristics in the one year previous to the lottery win. The coefficient $\beta_{3}$ stands for the difference-in-differences estimate for the baseline variable between current and future winners of the different prize amounts. 
Standard errors reported in parentheses. \\
*** significant at the 1 percent level,
** significant at the 5 percent level, and
* significant at the 10 percent level.
\end{tablenotes}
\end{threeparttable}
\end{center}

\begin{center}
 \begin{threeparttable}
 \setlength{\tabcolsep}{5mm}{}
 \linespread{0.85}
 \fontsize{8.5}{8.5pt}\selectfont
 \centering\footnotesize
 \caption{Descriptive Statistics for Lottery Winners (Re-weighted) and Population}
 \label{tab.descriptive.weighted}
 \begin{tabular}{@{}lccc@{}}
 \toprule
 & \begin{tabular}[c]{@{}c@{}}Lottery Winners\\(Re-weighted)\end{tabular} 
 & \begin{tabular}[c]{@{}c@{}}Population\end{tabular} 
 & \begin{tabular}[c]{@{}c@{}}Difference\end{tabular} \\
 \midrule \midrule

 \textit{Individual characteristics}  &  &  &  \\ 
 ~~Age & 31.357 & 31.355 & 0.002 \\
 & (7.895) & (7.896) & [0.013] \\
 ~~Living in urban area & 0.685 & 0.693 & -0.008*** \\
 & (0.465) & (0.461) & [0.001] \\
 ~~Female & 0.505 & 0.499 & 0.006*** \\
 & (0.500) & (0.500) & [0.001] \\
 ~~Married & 0.411 & 0.411 & 0.000 \\
 & (0.492) & (0.492) & [0.001] \\
 ~~Winner's Employment & 0.694 & 0.694 & 0.000 \\
 & (0.461) & (0.461) & [0.001] \\
 ~~Winner's Earnings  & 281.847 & 285.594 & -3.747*** \\
 & (452.819) & (546.489) & [0.728] \\
 ~~Winner's Income & 301.887 & 308.035 & -6.148*** \\
 & (509.486) & (656.776) & [0.822] \\
 ~~Winner's Assets  & 2,259.721 & 2,320.071 & -60.350*** \\
 & (9,923.573) & (13,292.058) & [16.055] \\
 ~~Winner's Liquid Assets  & 684.456 & 709.105 & -24.649*** \\
 & (5,889.549) & (7,938.590) & [9.532] \\
~~Winner's Savings & 262.720 & 292.212 & -29.492*** \\
 & (1,245.439) & (1,390.573) & [1.996] \\
~~Household Earnings  & 468.312 & 458.116 & 10.196*** \\
 & (735.580) & (869.859) & [1.182] \\
 ~~Household Income  & 502.935 & 497.178 & 5.757*** \\
 & (818.461) & (1,343.545) & [1.344] \\
 ~~Household Assets  & 4,033.504 & 4,165.627 & -132.123*** \\
 & (15,520.106) & (41,404.719) & [27.292] \\
 ~~Household Liquid Assets  & 1,144.448 & 1,209.224 & -64.776*** \\
 & (7,912.306) & (38,197.793) & [16.854] \\
 ~~Household Savings & 438.958 & 478.295 & -39.337*** \\
 & (1,824.363) & (2,440.002) & [2.951] \\
 \midrule
 \textit{Fertility variables} &  &  &  \\
 ~~Cumulative Number of Children & 0.834 & 0.817 & 0.017*** \\
 & (1.107) & (1.105) & [0.002] \\
 ~~Gave Birth in $s - 1$ & 0.035 & 0.034 & 0.001*** \\
 & (0.183) & (0.180) & [0.000] \\
 ~~Gave Birth in $s - 2$ & 0.035 & 0.034 & 0.001*** \\
 & (0.184) & (0.181) & [0.000] \\
 ~~Gave Birth in $s - 3$ & 0.037 & 0.035 & 0.002*** \\
 & (0.188) & (0.185) & [0.000] \\
 \midrule \midrule
 \# of Observations & 406,963 & 11,205,868 & \\
 \bottomrule
 \end{tabular}
 \begin{tablenotes}
 \fontsize{8}{8pt}\selectfont
Note: We utilize the all individuals aged 20-44 from 2007-2012 to construct population data. For each individual, we randomly assign one year between 2007-2012 as a placebo "winning year." We then use their individual characteristics from the year prior to this randomly assigned placebo winning year in our analysis. We use the post-stratification weighting technique and match the marital status, age, earnings, and asset stratifications for our lottery sample and the population. Urban areas refer to the 6 largest cities in Taiwan with special municipality status: Taipei City, New Taipei City, Taoyuan City, Taichung City, Tainan City, and Kaohsiung City. These cities have the largest populations in Taiwan. Employment is defined as having positive annual labor earnings. Annual earnings are defined as the sum of annual wage income, business income, and professional income. Annual income is defined as the sum of annual labor earnings plus other annual income sources like interest, rents, farming, pensions etc, excluding lottery winnings. Assets are defined as the sum of real estate value, financial assets, and stocks, minus mortgage debt. Liquid assets are defined as the sum of financial assets and stocks. All monetary values like earnings, income, assets and liquid assets are measured in thousand New Taiwan Dollars (NT\$)  and adjusted to 2016 NT\$ levels (1 NT\$ $\approx$ 0.033 US\$ in 2016). More details on the construction of asset data can be found in Appendix \ref{app: w_data}. Standard deviations are in parentheses, and standard errors are in brackets. *** significant at the 1 percent level, ** significant at the 5 percent level, and * significant at the 10 percent level.
 \end{tablenotes}
 \end{threeparttable}
\end{center}

\end{document}